\documentclass[aps,prd,showpacs,superscriptaddress,onecolumn,raggedfooter,raggedbottom]{revtex4}   

\usepackage{amssymb}
\usepackage{amsmath}
\usepackage{amsthm}
\usepackage{graphicx}
\usepackage{subfigure}
\usepackage{color}
\usepackage{mathrsfs} 
\usepackage{paralist}
\usepackage{comment}

\theoremstyle{definition} 
\newtheorem{defn}{Definition}

\usepackage{hyperref}
\hypersetup{hidelinks}
\usepackage{soul}

\usepackage[capitalize,nameinlink]{cleveref}
\crefname{section}{section}{sections}
\crefname{subsection}{subsection}{subsections}
\Crefname{section}{Section}{Sections}
\Crefname{subsection}{Subsection}{Subsections}
\Crefname{figure}{Figure}{Figures}
\crefformat{equation}{\textup{#2(#1)#3}}
\crefrangeformat{equation}{\textup{#3(#1)#4--#5(#2)#6}}
\crefmultiformat{equation}{\textup{#2(#1)#3}}{ and \textup{#2(#1)#3}}
{, \textup{#2(#1)#3}}{, and \textup{#2(#1)#3}}
\crefrangemultiformat{equation}{\textup{#3(#1)#4--#5(#2)#6}}%
{ and \textup{#3(#1)#4--#5(#2)#6}}{, \textup{#3(#1)#4--#5(#2)#6}}{, and \textup{#3(#1)#4--#5(#2)#6}}
\Crefformat{equation}{#2Equation~\textup{(#1)}#3}
\Crefrangeformat{equation}{Equations~\textup{#3(#1)#4--#5(#2)#6}}
\Crefmultiformat{equation}{Equations~\textup{#2(#1)#3}}{ and \textup{#2(#1)#3}}
{, \textup{#2(#1)#3}}{, and \textup{#2(#1)#3}}
\Crefrangemultiformat{equation}{Equations~\textup{#3(#1)#4--#5(#2)#6}}%
{ and \textup{#3(#1)#4--#5(#2)#6}}{, \textup{#3(#1)#4--#5(#2)#6}}{, and \textup{#3(#1)#4--#5(#2)#6}}
\crefdefaultlabelformat{#2\textup{#1}#3}

\begin{document}

\title{Bright Fractional Single and Multi-Solitons in a Prototypical Nonlinear Schr{\"o}dinger Paradigm: Existence, Stability and Dynamics}

\author{Robert J.\ Decker}
\affiliation{Mathematics Department, University of Hartford, 200 Bloomfield Ave., West Hartford, CT 06117, USA}

\author{A.\ Demirkaya}
\affiliation{Mathematics Department, University of Hartford, 200 Bloomfield Ave., West Hartford, CT 06117, USA}

\author{T.J. Alexander}
\affiliation{Institute of Photonics and Optical Science (IPOS), School of Physics, The University of Sydney, NSW 2006, Australia}

\author{P.~G.\ Kevrekidis}
\affiliation{Department of Mathematics and Statistics, University of Massachusetts,Amherst, MA 01003-4515, USA}

\begin{abstract}
In the present work we explore features of single and pairs of solitary
waves in a fractional variant of the nonlinear Schr{\"o}dinger equation.
Motivated by the recent experimental realization of arbitrary fractional exponents,
upon quantifying the tail properties of such coherent structures,
we detail their destabilization  when the fractional
exponent $\alpha$ acquires values $\alpha<1$ and showcase how the relevant
destabilization is associated with collapse type phenomena. We then turn to
in- and out-of-phase pairs of such waveforms and illustrate how they
generically exist for arbitrary $\alpha$ when we cross the harmonic limit,
i.e., for $\alpha>2$. Importantly, we use the parameter $\alpha$ as a ``bifurcation
parameter'' in order to connect the harmonic ($\alpha=2$) and biharmonic
($\alpha=4$) limits. Remarkably, not only do we retrieve the instability of
all solitonic pairs in the biharmonic case, but showcase a 
stabilization feature of particular branches of such multipulses
that is {\it unique} to the fractional case and does not arise ---to our knowledge---
for integer multi-pulse settings. We explain systematically this
stabilization via spectral analysis and expand upon the implications of our results for the potential
observability of fractional multipulse solitary waves.
\end{abstract}

\maketitle

\section{Introduction}
Over the past decade, fractional-calculus--based models have attracted 
a significant amount of attention, arguably sparked by a wide spectrum of potential applications. Representative examples include 
epidemiological modeling~\cite{qureshi2020real},
fractional diffusion in biological systems~\cite{IonescuCNSNS2017},  the spread of computer viruses~\cite{singh2018fractional,azam2020numerical}, notably ---especially for our purposes--- optical systems~\cite{malomed}, as well as nonlinear wave dynamics~\cite{cuevas}, and economic modeling~\cite{ming2019application}. This
ever expanding stream of activity  has led to several comprehensive reviews and monographs; see, e.g.,~\cite{podlubny1999fractional,Samko,Mihalache2021}.

In parallel to application-driven studies, substantial effort has been devoted to the mathematical structure and associated understanding of fractional operators and their properties~\cite{yavuz2020comparing,saad2018new}. Nonetheless, many fundamental issues remain unresolved~\cite{ortigueira2021two,ortigueira2021bilateral}, in part due to the coexistence of multiple definitions of fractional derivatives. Among these, Riesz fractional derivatives stand out as particularly appealing for physical applications~\cite{muslih2010riesz}, as they arise naturally as continuum limits of lattice systems with long-range interactions~\cite{tarasov2006fractional}. This connection provides an important bridge between microscopic models and effective fractional-order continuum descriptions.

Despite the ever-expanding richness of the theoretical/mathematical literature, experimental confirmations/realizations of fractional-order wave dynamics have  been relatively scarce. A notable exception has recently emerged in nonlinear optics, where remarkable progress has been made in engineering dispersion landscapes and controlling the effective order of dispersion~\cite{BlancoRedondoNC2016,RungeNP2020}. 
Indeed, these works produced the seminal achievement of the so-called pure 
quartic solitons and sparked an extensive interest in their analysis,
and experimental (as well as theoretical) generalizations~\cite{deSterkeRungeHudsonBlancoRedondoAPL2021PQS,ParkerAcevesPhysD2021Multipulse,DeckerDemirkayaMantonKevrekidisJPA2020BiharmonicPhi4,TsoliasDeckerDemirkayaAlexanderParkerKevrekidisCNSNS2023NLSMixedDispersion,TamOL2019,TamPRA2020,BandaraPRA2021}.
More recent efforts have extended to higher-order settings~\cite{QiangAlexanderdeSterkePRA2022SixthOrder,deSterkeBlancoRedondoOptCom2023EvenOrder,WidjajaQiangSkeltonNatCommun2025Universality}. It is worthwhile to
note in passing
that considerably less effort has been dedicated to dark solitonic
structures, even though these may be interesting in their own 
right~\cite{Alexander:22}.

The above developments have provided a concrete physical pathway connecting higher-order dispersion models with experimentally observable solitary waves. Importantly, they 
have also motivated the exploration of fractional dispersion as a tool enabling
the homotopic interpolation between harmonic and biharmonic (i.e., quartic) limits. This naturally leads to the question of how solitary-wave (and multi-solitary-wave) properties evolve as the dispersion order is varied continuously rather than 
discretely between these two limits, a question that was recently touched
upon for kinks in a real field-theoretic model~\cite{DECKER2025100051}.
At the early stages of fractional model studies in optics, 
the fractional Schr\"odinger equation was proposed as an optical propagation model in~\cite{Longhi:15} and subsequently implemented experimentally in the linear regime~\cite{malomed}. More recently, a series of experiments have demonstrated direct realizations of fractional dispersion and fractional evolution in optical platforms, including temporal-domain implementations of the fractional Schr\"odinger equation~\cite{LiuNatCommun2023FSE}, as well as ---most importantly for our
purposes--- nonlinear propagation governed by a fractional derivative in the 
pioneering work of~\cite{HoangNatCommun2025FractionalDerivative}. 
The latest variant of such experimental realizations involves a 
parametric variation of the dispersion relation in Fourier space
(albeit not in the fractional exponent $\alpha$ as is considered
herein) from the $\alpha=1$ so-called Hilbert variant of the 
nonlinear Schr{\"o}dinger (NLS) equation to the so-called Nyquist
variant thereof in~\cite{martijn25}.

The particularly rich phenomenology of higher-order dispersion systems originates in the structure of their solitary-wave tails. In the pure-quartic case, for example, the biharmonic operator induces oscillatorily decaying tails associated with complex spatial eigenvalues~\cite{DeckerJPA2020}. This mechanism underlies the emergence of multiple equilibria, bound states, and self-similar interaction dynamics. In contrast, the standard harmonic NLS equation~\cite{KIVSHAR1995353}---and its real-field counterpart, the $\phi^4$ model~\cite{p4book}---possess only real spatial eigenvalues, leading to monotonic exponential interactions and the absence of solitonic bound states.

Our aim herein is to provide a connection between these two limits for the
realm of bright solitary waves and multi-solitary waves in the experimentally
tractable model of the fractional nonlinear Schr{\"o}dinger equation~\cite{HoangNatCommun2025FractionalDerivative}. 
Extending the above-mentioned theoretical and experimental analysis, our
aim is to examine the continuation of single solitary
waves for arbitrary exponents $\alpha$, associated with the Riesz spatial
derivative. For all $\alpha$'s, we examine the tails of the solitary waves, and of
the multi-solitary-wave structures. 
In connection to stability, we find a threshold of $\alpha=1$ (when $\alpha=2$ denotes
the Laplacian limit), where the stability of the solitary waves changes.
The relevant destabilization, upon crossing this threshold, is found to be
connected to collapse type dynamics~\cite{Sulem,fibich2015} which we monitor
and connect to the associated linearization spectrum of the solitary wave.
Then, while below the sub-harmonic limit, no multi-solitary-wave steady state
exists, this drastically changes as soon as one goes above the Laplacian
($\alpha=2$) case, with one such stationary waveform immediately forming,
infinitesimally away from that limit. Using $\alpha$ as a continuation parameter,
we identify all the multi-soliton waveforms for different $\alpha$ and
how they terminate, as well as how they connect the limit of no-multisolitons
($\alpha=2$) with that of infinitely many such $(\alpha=4)$. This continuation
presents remarkable features, since while it has been shown
that {\it all} bright multi-pulses are unstable in the biharmonic ($\alpha=4$)
limit~\cite{ParkerAcevesPhysD2021Multipulse}, yet windows of stability are found
and explained within the fractional model for the even (but not for the
odd) multipulse solution branches.

Our presentation is structured as follows. 
In the short section II, we provide the setup of the mathematical problem.
Section III discusses the single soliton solutions~\footnote{We should clarify henceforth that we will use the term ``soliton'' in a somewhat loose fashion, meaning solitary waves, as is often done in the nonlinear waves community. This is with the understanding that merely the case of $\alpha=2$ corresponds to the integrable
focusing NLS where true solitons exist, while generically for all other values
of $\alpha$, these are solitary waves of a non-integrable model.}, 
as concerns their tails, their stability and associated oscillatory or unstable
dynamics (on the two sides of $\alpha=1$). Section IV is dedicated to
two-solitary-wave solutions, building the branches' bifurcation diagram, analyzing
their stability as a function of $\alpha$ and exploring their different type
of unstable dynamics (both positional shift-inducing instabilities and 
symmetry-breaking instabilities). Finally, section V summarizes our findings
and presents our conclusions, as well as a number of directions for future
study.

\section{Problem Setup and Numerical Methods}
\subsection{Problem Setup}

We consider the fractional focusing NLS (FF-NLS) equation with cubic nonlinearity in the form
\begin{equation}
iu_{t}=D^{\alpha }u+2u|u|^{2}.  \label{eq:nls1}
\end{equation}%
Notice that while the model is often written with the opposite sign, here
we leverage the time-reversal symmetry to express it in the form of Eq.~(\ref{eq:nls1}).
When $\alpha=2$, and hence $D^\alpha u=D^2u=u_{xx}$, we get the standard NLS 
equation~\cite{Sulem,fibich2015}. In the present work we will consider $0<\alpha \le 4$ in which case $D^\alpha$ is defined as the Riesz (fractional) derivative in the Fourier sense, given by
\begin{equation}
D^{\alpha }f=F^{-1}[(-|k|^{\alpha })\widehat{f}(k)].  
\label{eq:riesz}
\end{equation}%
Here, the hat symbol denotes the Fourier transform, $F^{-1}$ its inverse, and $k$ the Fourier variable. Note that $D^\alpha$ can also be written as $-(-\Delta)^{\alpha/2}$ where $\Delta$ is the Laplacian operator.

If we now assume a solution to Eq. (\ref{eq:nls1}) of the form $u(x,t)=\exp(-i \omega t)\phi(x)$, 
where $\phi(x)$ is real valued, we get
\begin{equation}
D^{\alpha }\phi+2\phi^{3}-\omega \phi =0 \label{eq:nlsSteady}
\end{equation}%
which is the stationary FF-NLS equation. For the rest of the paper we will use $\omega=1$.

Accordingly, linearizing via
\begin{equation}
    u(x,t)=e^{-it}\left[\phi(x)+\epsilon(p(x,t)+iq(x,t))\right],
    \label{linearize}
\end{equation}
yields at $O(\epsilon)$
\begin{align}
p_t &= D^\alpha q + (2\phi^2-1)q, \\
q_t &= -D^\alpha p - (6\phi^2-1)p.
\end{align}

Seeking normal modes $p=P e^{\lambda t}$, $q=Q e^{\lambda t}$ gives
\begin{equation}
\lambda
\begin{pmatrix} P \\ Q \end{pmatrix}
=
\begin{pmatrix}
0 & L_+ \\
- L_- & 0
\end{pmatrix}
\begin{pmatrix} P \\ Q \end{pmatrix},
\label{linearize2}
\end{equation}
where
\begin{eqnarray}
L_+ = D^\alpha + (2\phi^2-1), 
\label{L+}
\qquad
L_- = D^\alpha + (6\phi^2-1).
\label{L-}
\end{eqnarray}
In what follows, we will see stationary solutions of the FF-NLS equation
satisfying Eq.~(\ref{eq:nlsSteady}), for which we will subsequently analyze
the linearization of Eq.~(\ref{linearize2}). Finally, when we find interesting
features within this linearization, such as internal breathing modes, or
real eigenvalues associated with instabilities, we will seek to explore the
dynamical implications of such modes, via direct numerical simulations
of Eq.~(\ref{eq:nls1}). We will start our analysis with single-hump
solitary wave solutions, and subsequently turn to the more complex
multi-soliton solutions.

\subsection{Numerical Methods}\label{numericalMethods}

We can discretize Eq. (\ref{eq:nlsSteady}) by first discretizing the continuous $x$ variable on an interval $[-L,L]$ using $N+1$ equally spaced intervals to get the discrete $x_i$. Base values of at least L=80 and N=1200 were used, with larger values required in some cases. Then with $\phi_i=\phi(x_i)$ we can approximate the Riesz derivative of order $\alpha$ using the Fourier definition as described in Eq. (\ref{eq:riesz}) with the FFT replacing the Fourier Transform.

Steady states can then be found using the Matlab `fsolve' command on the left side of the discrete version of Eq. (\ref{eq:nlsSteady}), along with a suitable initializer. The single soliton initializers for all $\alpha$ were the exact solution for the case $\alpha=2$ given by $\phi_{0}(x)=\operatorname{sech}(x)$ for our chosen $\omega=1$. For two-soliton pairs (in-phase) we can use the function
\begin{equation}
\phi(x) = \phi_{0}(x+\delta)+\phi_{0}(x-\delta),\label{twoSolInit}
\end{equation}%
as an initial guess for the stationary solution (for out-of-phase include negative sign on first term), where $2\delta$ is the separation distance between the two solitons. For a given $\alpha$ (in the range $2<\alpha<4$), there will be a finite number of values of $\delta$ which result in steady states. We can find those $\delta$ values and the corresponding steady states using a simple numerical continuation, starting at $\alpha=4$ (and $\delta$ determined via search) and then reducing $\alpha$ by a small increment and using the previous initializer as the initializer for the new $\alpha$-value. This results in a new $\delta$-value and then we can continue (near certain bifurcation points this may break down, and then a new search gets the process back on track). 

To determine stability of the various steady states, we created a fractional differentiation matrix
$D^\alpha$, based on Eq. ($\ref{eq:riesz}$), with discrete Fourier Transformation matrices replacing the continuous operators. Stability can then be determined by using the Matlab `eig' command to find the eigenvalues of the stability
matrix
$\begin{bmatrix}
0 & L_+ \\ 
-L_- & 0%
\end{bmatrix}$
arising in Equation (\ref{linearize2}). For PDE simulations we discretize the right side of Equation (\ref{eq:nls1}) as before, resulting in a system of N ODEs, which can be numerically solved using the Matlab routines `ode45' or `ode23'.

\section{Single Soliton Solutions}
\subsection{Tail Behavior}
It should be noted before we provide numerical details that the existence
problem of the fractional NLS has been explored in numerous publications
in the mathematical literature including the existence problem
in~\cite{Secchi2013} (through variational methods) and in a more general
setting in~\cite{DipierroSoaveValdinoci2017}, as well as the uniqueness
problem in the key work of~\cite{FrankLenzmann2013}. Similarly, 
numerical efforts have focused on the solitary waves and dynamics of
the FF-NLS; see, e.g.~\cite{KleinSparberMarkowich2014} for a representative
example, as well as the recent review of~\cite{malomed2}, as well as the book~\cite{cuevas}.
While the properties of the fractional linear operator dictating the tails, as well
as the potential for collapse have been explored in works such as the above, here we
complement these demonstrating a number of numerical case examples for the tail and showcasing
the relevant stability eigenvalues and their dependence on the fractional
exponent. These illustrations pave the way for the study of solitonic ``molecules'', i.e., of
pairs of solitary waves which constitute the focus of the present study.

\begin{figure}[h]
\begin{center}
\includegraphics[width=3.5cm]{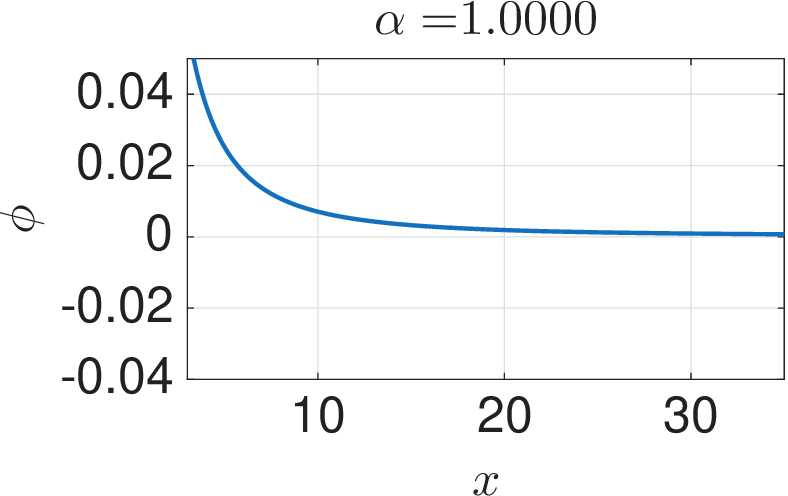} 
\includegraphics[width=3.5cm]{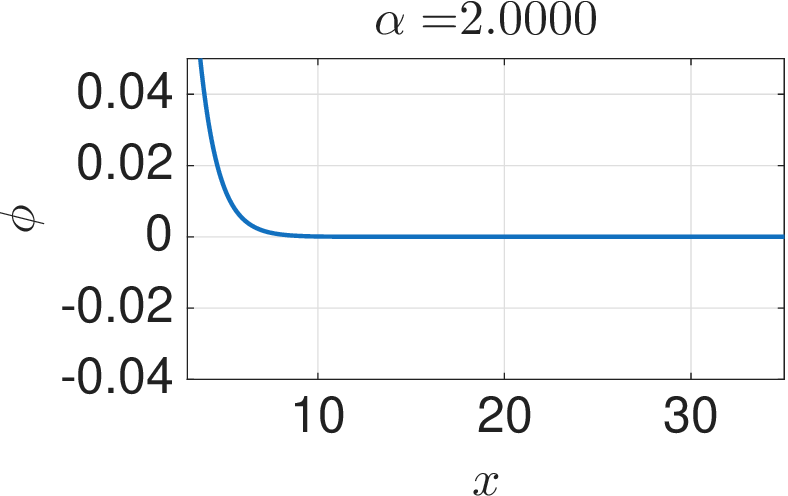} 
\includegraphics[width=3.5cm]{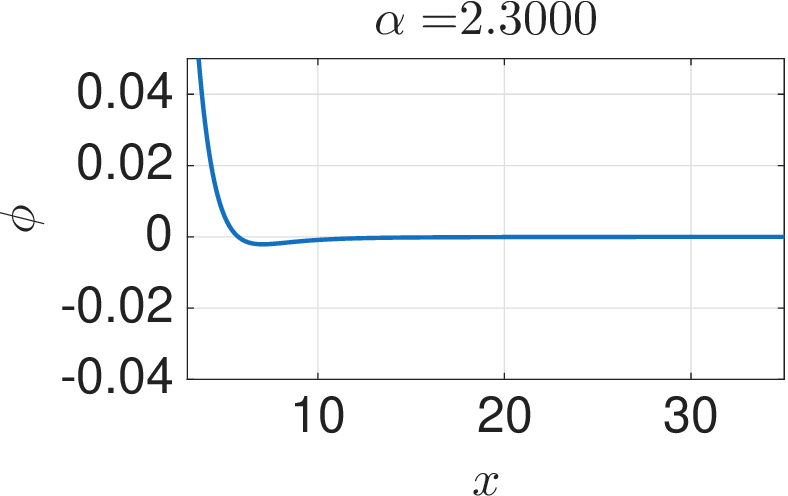} 
\includegraphics[width=3.5cm]{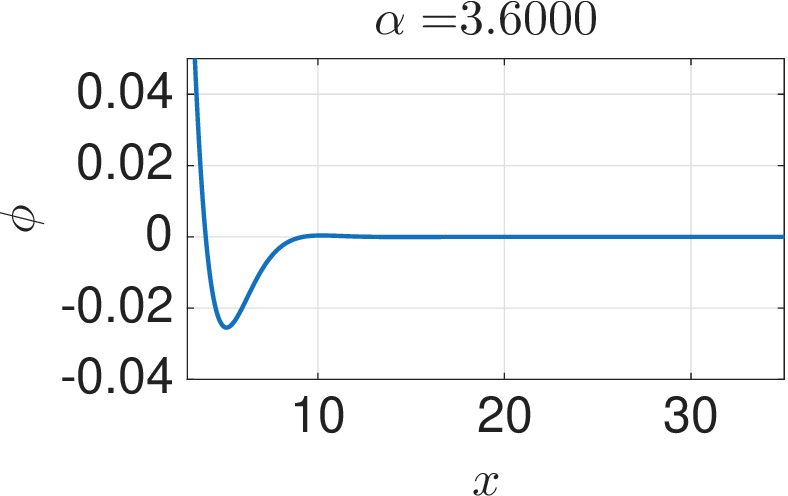} 
\includegraphics[width=3.5cm]{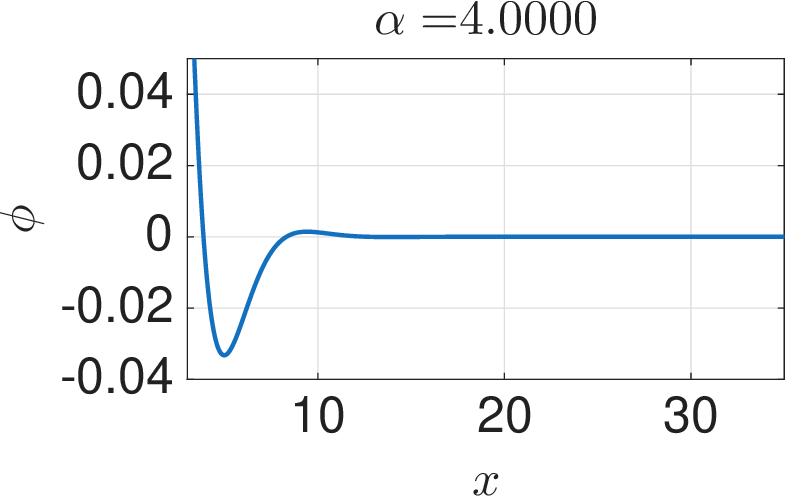} 
\end{center}
\begin{center}
\includegraphics[width=3.5cm]{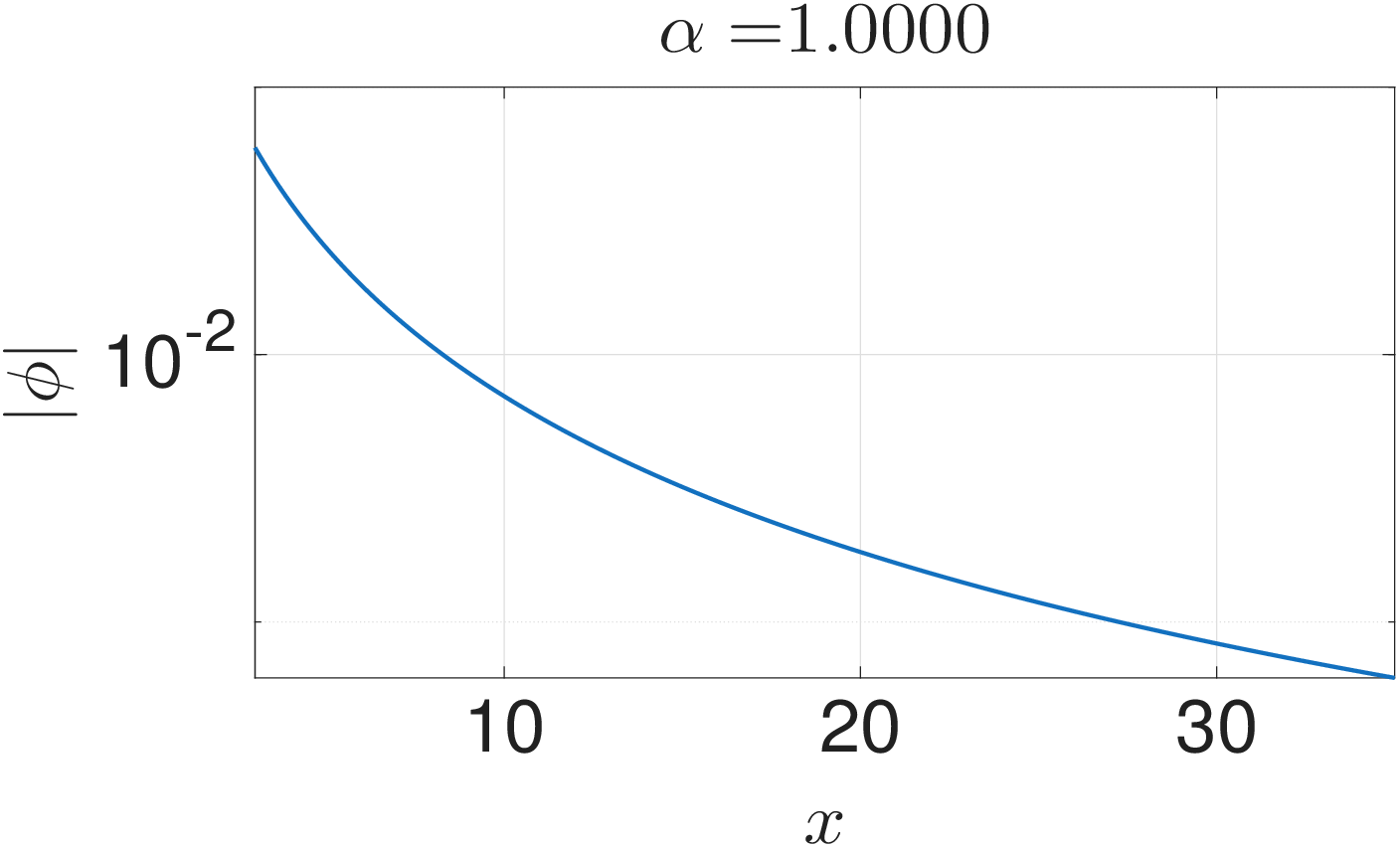} 
\includegraphics[width=3.5cm]{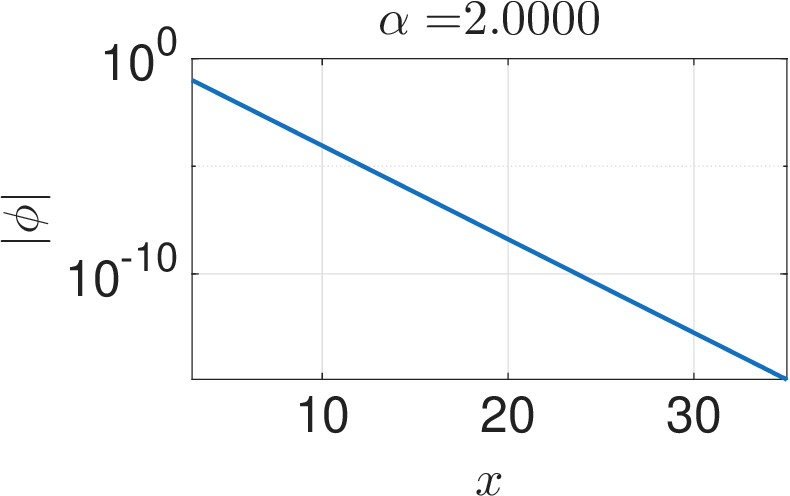} 
\includegraphics[width=3.5cm]{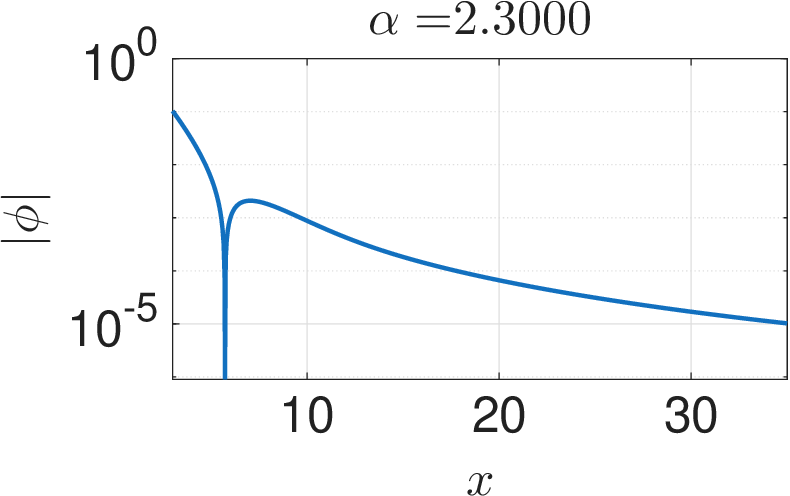} 
\includegraphics[width=3.5cm]{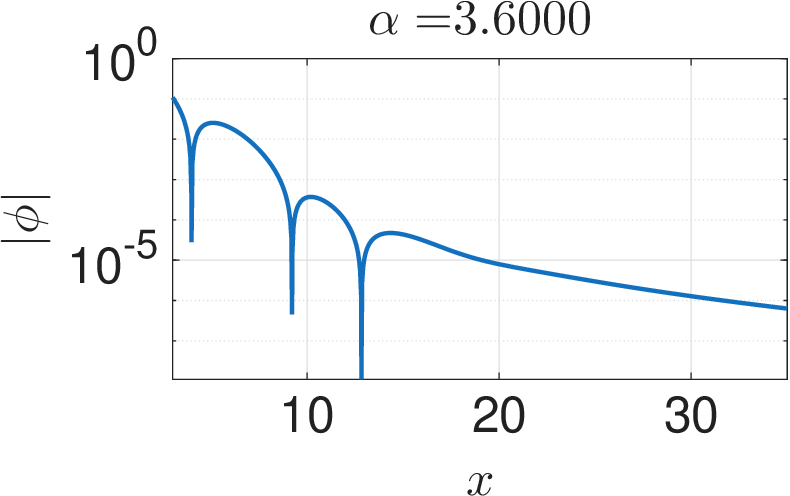} 
\includegraphics[width=3.5cm]{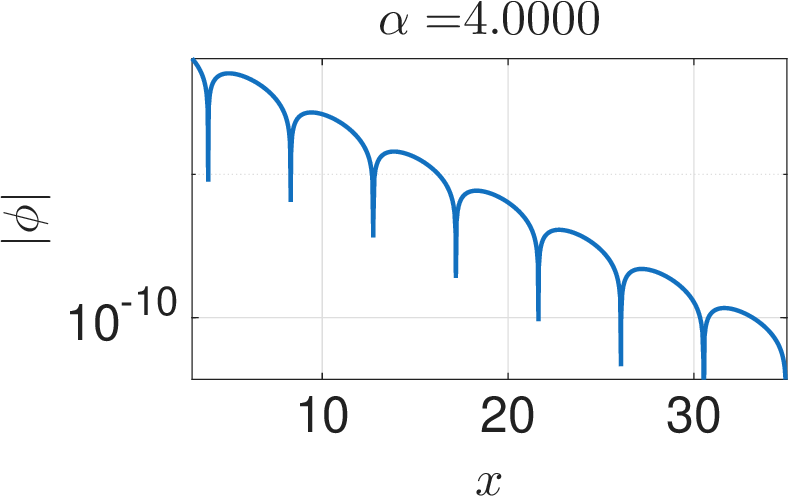} 
\end{center}
\caption{Right tails of the solitary wave solutions for various values of $\alpha$
in a linear scale (top) and a semilog scale (bottom).}
\label{fig_solitonShape}
\end{figure}

In our setting, for $0<\alpha \le 4$ we have numerically identified real-valued single soliton solutions to Eq. (\ref{eq:nlsSteady}). For $0<\alpha \le 2$, these solutions are positive for all $x$ and the tails are monotonically decreasing, so there are no zero crossings. For $2<\alpha < 4$ there is at least one zero crossing and the tail approaches zero from below, in a way reminiscent also of corresponding 
results in~\cite{atanas2}. 
As $\alpha \xrightarrow{}4$ the number of zero crossings increases without bound, and at $\alpha=4$ there are infinitely many crossings. Additionally, in the cases where there is more than one zero crossing, the oscillations can be very accurately approximated by an exponentially damped sine (or cosine) function. For $0<\alpha<4, \alpha \neq 2$ the far tails can be approximated by a power law function of the form $a / x^b$, where $b=\alpha+1$, in line with the
corresponding theoretical prediction~\cite{Secchi2013}; see the panels of 
Figure \ref{fig_solitonShape}.

\begin{figure}[h]
\begin{center} 
\includegraphics[width=6.9cm]{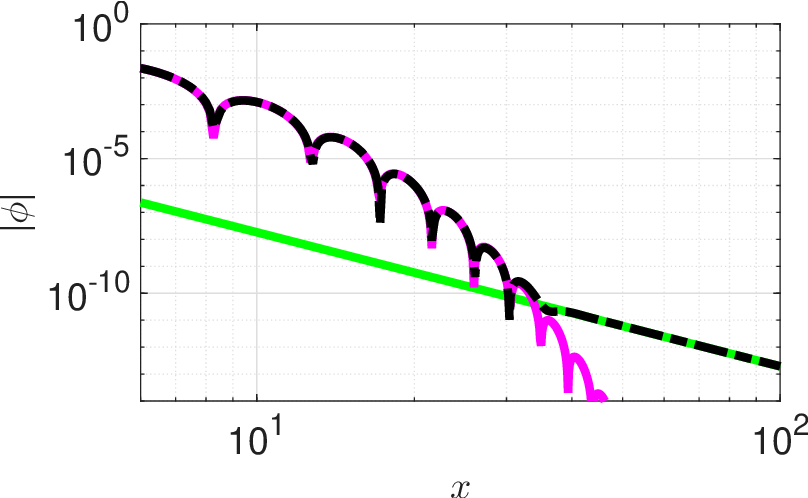} 
\includegraphics[width=6.9cm]{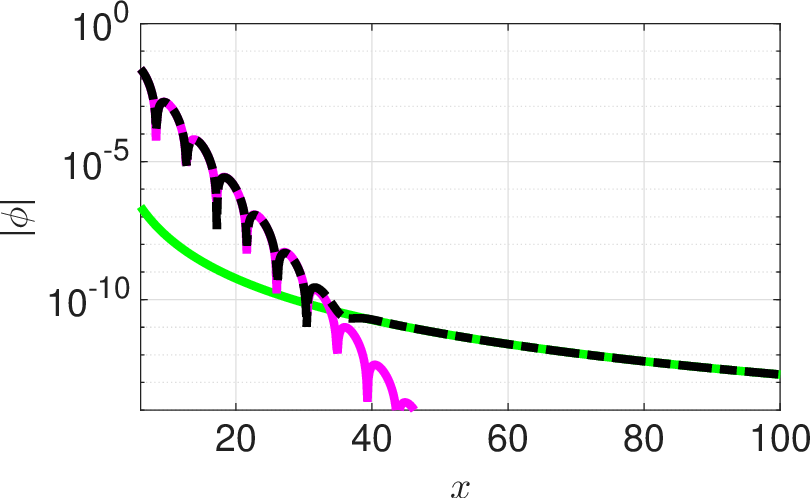} 
\end{center}
\caption{Tails of a stationary solution, and fitted curves for $\alpha=3.99995$. The stationary solution is a black dash-dot line, the oscillating part of the tail 
(the ``near tail'') is approximated by $\phi(x)=1.618\,{\mathrm{e}}^{-0.7085\,x} \,\sin \left(0.7066\,x+0.4376\right)$ (magenta curve) and the non-oscillating (power law)
part (the ``far tail'') is approximated by $\phi(x)=\frac{0.001886}{x^{4.998} }$ (green curve). The left panel is presented in a loglog plot and the right one in a  semilog plot.}
\label{fig_tails_alpha_399995}
\end{figure}

Further detail of the relevant tails is provided in Figure \ref{fig_tails_alpha_399995},
containing two views of the right tail of a single solitary wave for $\alpha=3.99995$ where there are seven zero-crossings of the $x$ axis, along with curves fitted to the numerically derived stationary state for both the oscillating part (what we
refer to as ``near tail'') and the power-law  decay part (what we refer
to as ``far tail'').

\subsection{Stability}
\begin{figure}[h]
\begin{center}
\subfigure[]{\includegraphics[width=4.1cm]{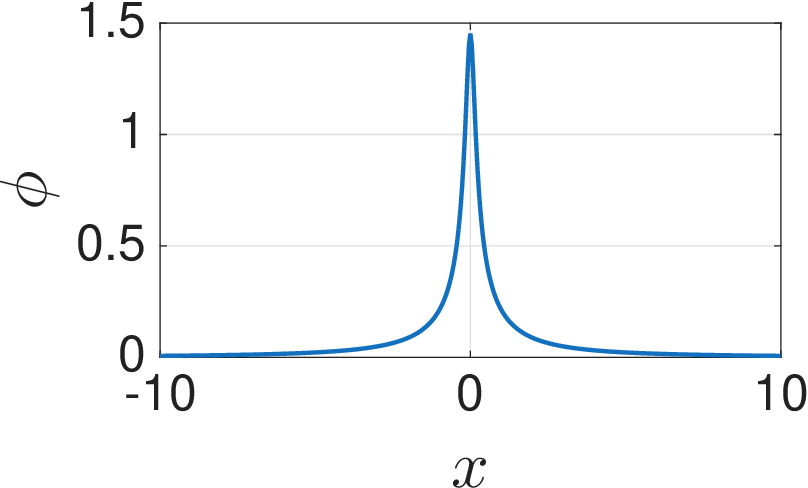} }
\subfigure[]{\includegraphics[width=4.1cm,height=2.38cm]{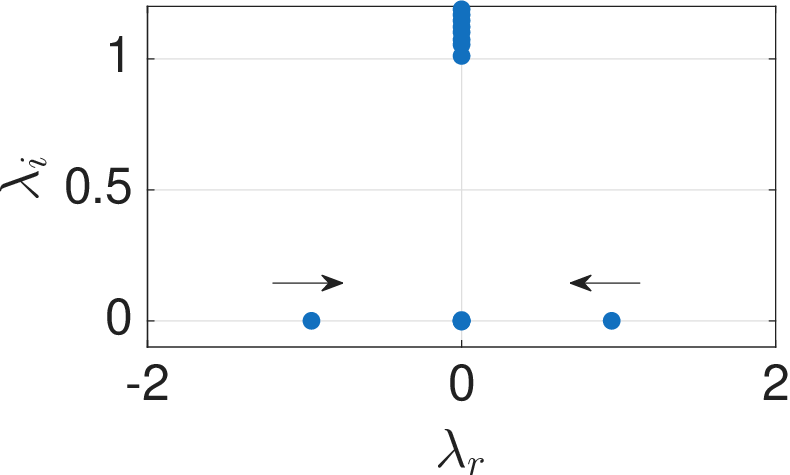} }
\subfigure[]{\includegraphics[width=4.1cm]{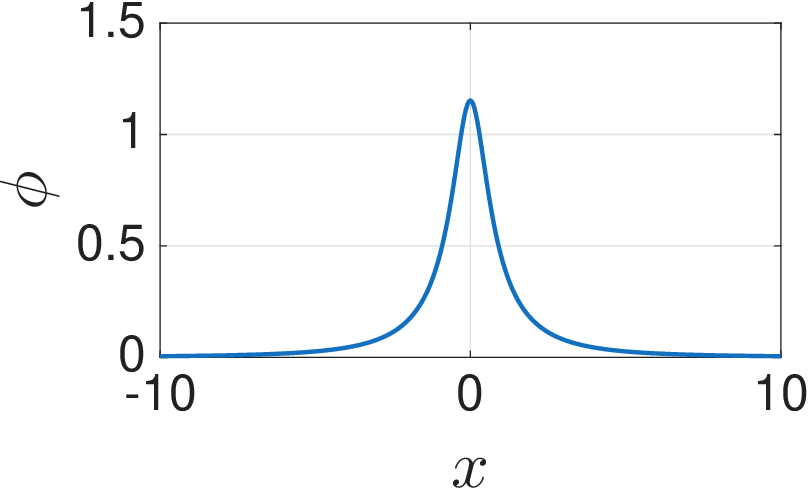} }
\subfigure[]{\includegraphics[width=4.1cm,height=2.38cm]{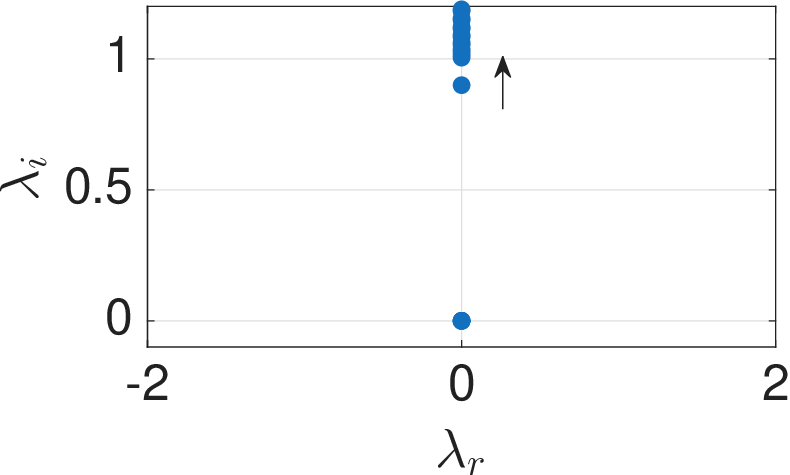} }
\end{center}
\caption{Single soliton plots and spectral plots: $\alpha=0.9$ for panels (a) and (b) and $\alpha=1.3$ for panels (c) and (d). Eigenvalue plots include only the positive imaginary axis, as the negative part is symmetric about the real axis. Isolated non-zero real valued eigenvalues in (b) are given by $\lambda=\pm 0.9593$ and isolated non-zero imaginary eigenvalues in (d) are $\lambda=\pm 0.8977i$. Arrows indicate the motion of the eigenvalues   as $\alpha$ increases. The eigenvalues that are on the $x$-axis for $\alpha=0.9$ pass through the origin and switch to the imaginary axis at $\alpha=1.3$; the switch occurs at $\alpha=1$ (as indicated in Figure \ref{alphaSquared}).}
\label{fig_spectralOneSoliton}
\end{figure}

\begin{figure}[h]
\begin{center}
\includegraphics[width=8cm]{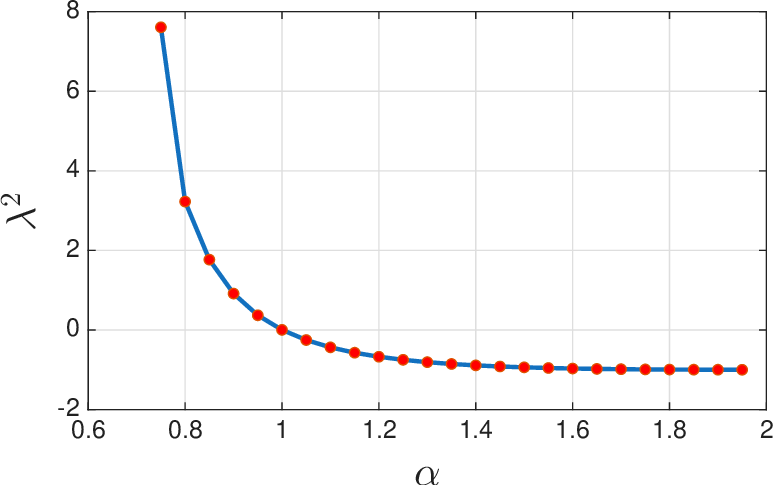} 
\end{center}
\caption{$\lambda^2$ as a function of $\alpha$ for the isolated (i.e., point
spectrum) eigenvalue pair $\pm\lambda$ that depends on $\alpha$, for the single soliton state (as shown in Figure \ref{fig_spectralOneSoliton}). We see that the pair of real eigenvalues in Figure \ref{fig_spectralOneSoliton} first moves toward the origin, then moves away from the origin on the imaginary axis. Thus, $\lambda^2$ decreases towards zero for real values ($\alpha \le 1$), then smoothly transition to negative (and decreasing) for imaginary values ($\alpha>1$). This is in line with the expected fundamental
solution instability within this model
for $\alpha < d$ (where $d$ is the dimension of the operator).}
\label{alphaSquared}
\end{figure}

Using the linearization formulation of Eqs.~(\ref{linearize})-(\ref{linearize2}),
in Figure \ref{fig_spectralOneSoliton}
we show soliton profiles and spectral plots $(\lambda_r,\lambda_i)$ for 
the eigenvalues $\lambda=\lambda_r + i \lambda_i$ 
corresponding to the relevant steady state for
$\alpha=0.9$ and $\alpha=1.3$. 
Recall that the expectation here is that for the formulation of Eq.~(\ref{eq:nls1}),
the critical case separating stability from instability and enabling the 
potential for collapse is $\alpha=d$, where $d$ denotes the dimensionality of
the relevant operator~\cite{KleinSparberMarkowich2014}. Indeed, instability 
is expected for $\alpha<d$, while stability emerges for the cases of
$\alpha>d$.
For $\alpha=0.9$ there indeed exists a pair of real eigenvalues at $\lambda=\pm 0.959$ reflecting the instability. This is found to be the case for all $0<\alpha<1$, and as $\alpha$ approaches $\alpha=1$ this real pair approaches the origin. As $\alpha$ passes through $\alpha=1$ the real pair of eigenvalues switches to the imaginary axis (transitioning the relevant state from instability to stability) and moves toward $\pm i$ as $\alpha \xrightarrow{}2$. The relevant eigenvalue variation with $\alpha$
has been indicated by an arrow in the figure panels.
In Figure \ref{alphaSquared} we showcase this eigenvalue variation with a plot of $\lambda^2$ as a function of $\alpha$ for $0<\alpha\le 2$. When $\lambda^2<0$ the eigenvalue is imaginary ($\alpha \geq 2$) and the system is spectrally stable, 
and when $\lambda^2>0$ it is real. For $2<\alpha \le 4$, $\lambda^2$ remains negative (not shown), establishing the spectral stability in that case as well.

\subsection{Single Soliton dynamics}
\label{sub:dynamics1}
\begin{figure}[h]
\begin{center}
\subfigure[]{\includegraphics[width=3.4cm]{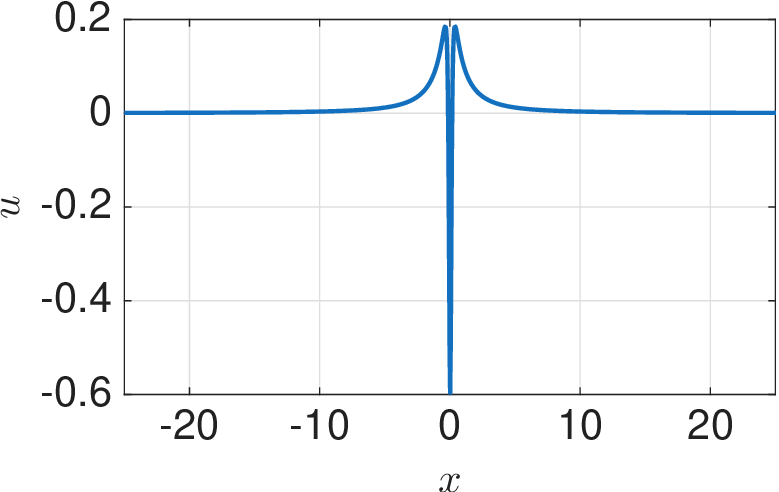} }
\subfigure[]{\includegraphics[width=3.4cm]
{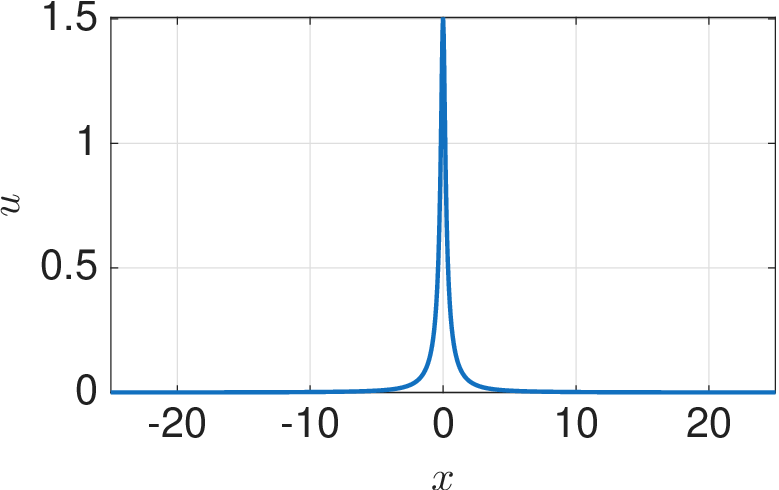} }
\subfigure[]{\includegraphics[width=3.6cm]{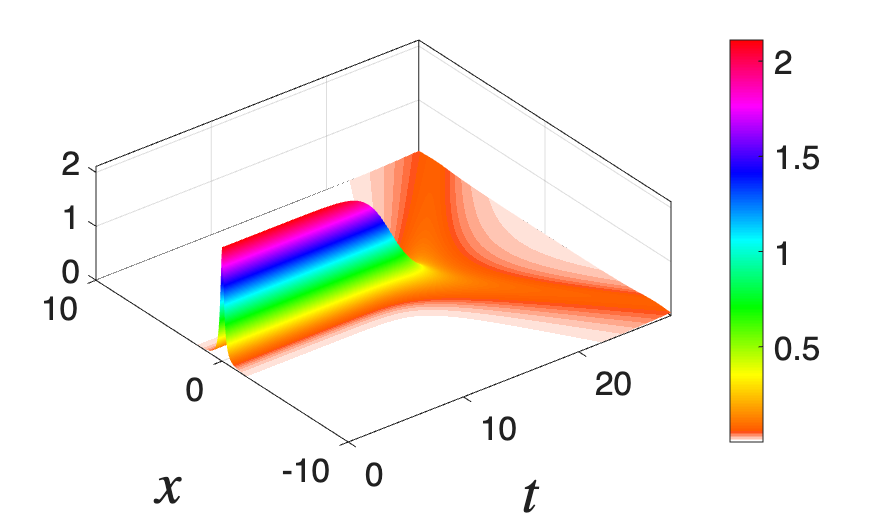}}
\subfigure[]{\includegraphics[width=3.6cm]{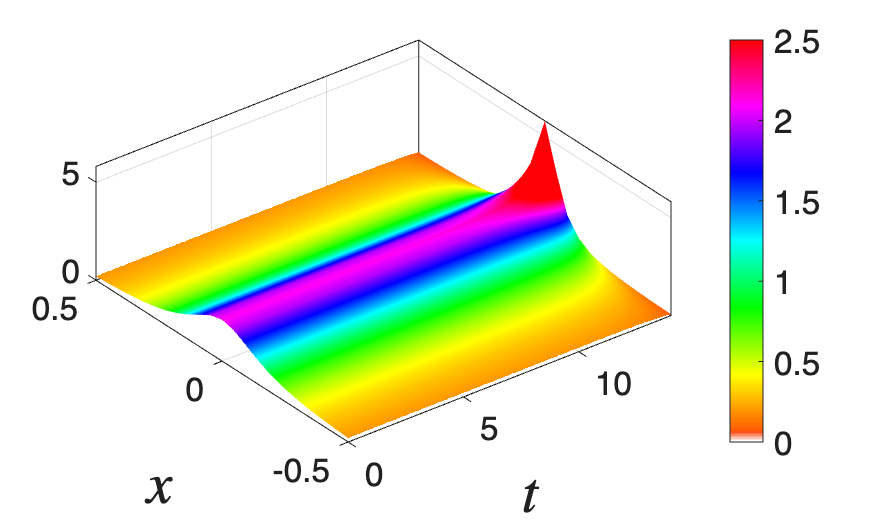}}
\subfigure[]{\includegraphics[width=3.3cm]{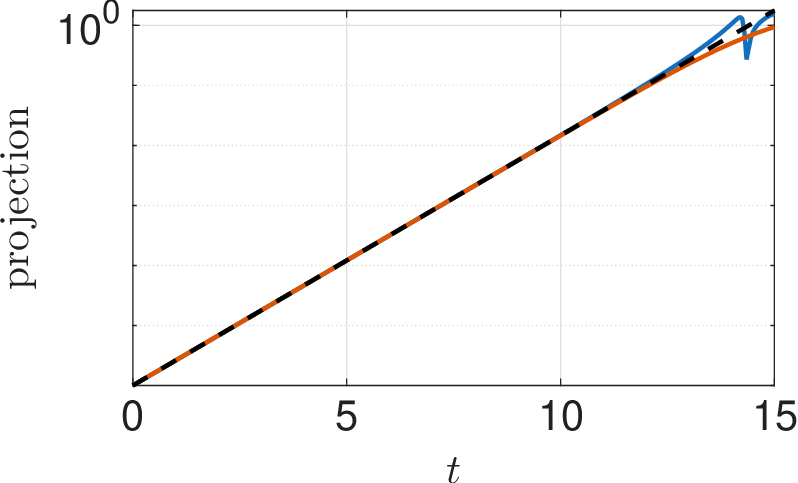}}
\end{center}
\begin{center}
\subfigure[]{\includegraphics[width=3.4cm]{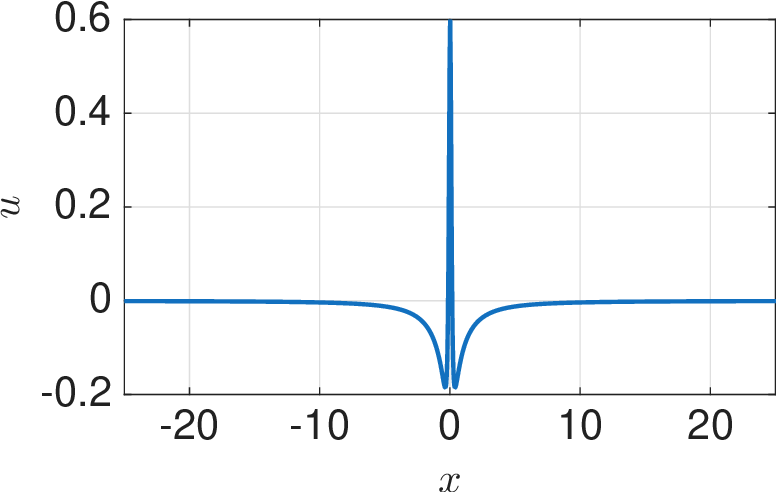} }
\subfigure[]{\includegraphics[width=3.4cm]
{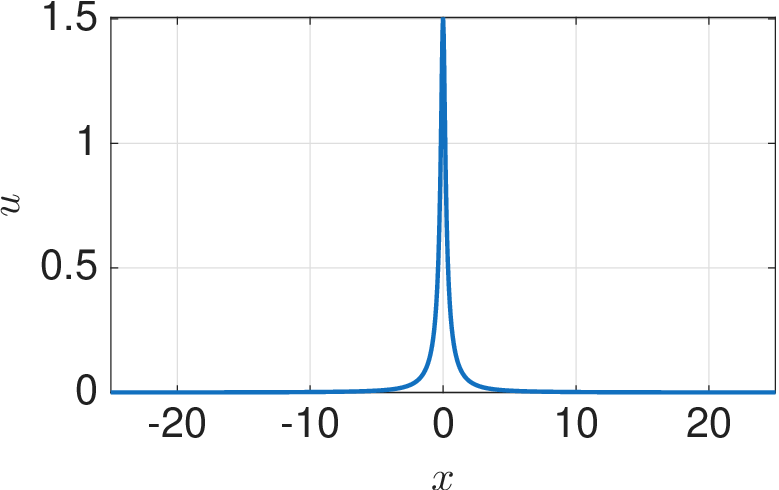} }
\subfigure[]{\includegraphics[width=3.6cm]{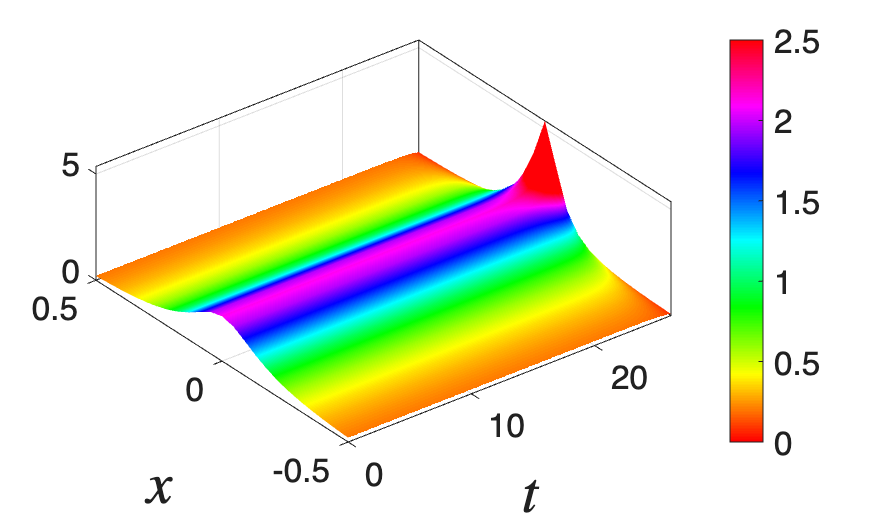}}
\subfigure[]{\includegraphics[width=3.6cm]{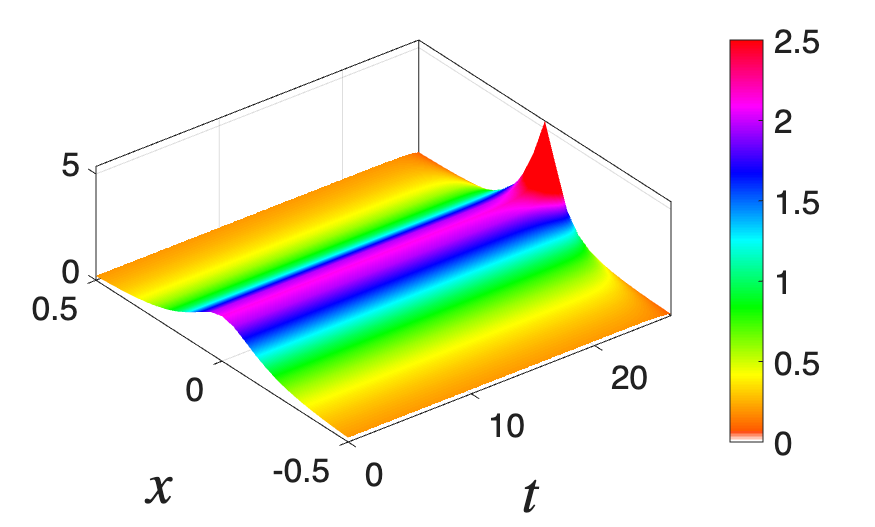}}
\subfigure[]{\includegraphics[width=3.3cm]{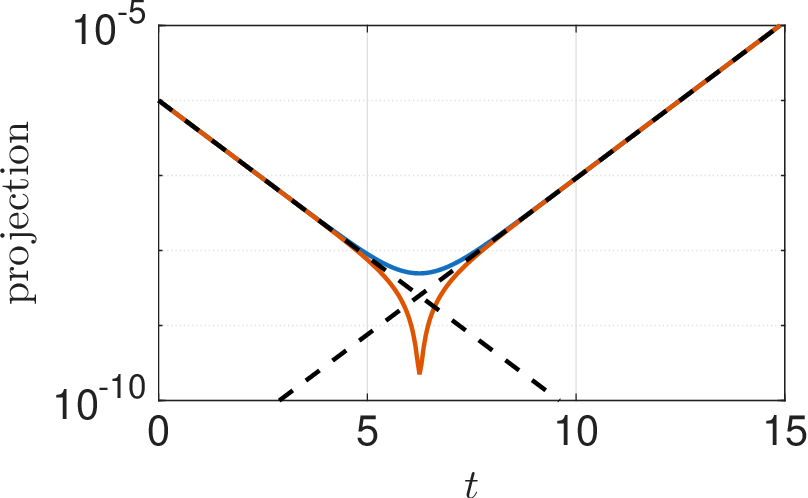}}
\end{center}
\caption{Eigenvectors and dynamical evolution of a single soliton for $\alpha=0.9$ and real eigenvalues $\lambda=\pm0.956$; top row is $\lambda=0.956$, bottom row $\lambda=-0.956$.  Panels (a), (b), (f), (g) are real part of eigenvector (left) and imaginary part of eigenvector (right).  Panels (c)-(e) and (h)-(j) are the results of a dynamic simulation using as initial condition ``steady state + $\varepsilon \times$eigenvector'' for the presented eigenvector. The  surface plots of $|u(x,t)|$ 
for $\varepsilon=10^{-6}$ (panels (c) and (h)) and  for $\varepsilon=-10^{-6}$ (panels (d) and (i)) are shown, as well as the projection of $u(x,t)-u(x,0)\exp(it)$ onto the given eigenvector (panels (d) and (j)) for both $\varepsilon=10^{-6}$ (in blue) and $\varepsilon=-10^{-6}$ (in red). }
\label{alpha0.9_1}
\end{figure}

We now look deeper into the stability/instability of single solitons indicated in the previous subsection. In Figure  \ref{alpha0.9_1} we show the dynamics that result when we use the steady state perturbed by a small multiple ($\varepsilon$) of the eigenvectors corresponding to the positive (top row (a)-(e)) and negative (bottom row (f)-(j) ) eigenvalues. In this figure we show the eigenvectors (i.e., their real and imaginary parts), surface plots of simulations for perturbations of $\varepsilon=10^{-6}$ and $\varepsilon=-10^{-6}$, and a projection onto the eigenvector for both the positive and negative perturbations.
For this case ($\alpha=0.9)$ the perturbed eigenstate either ends up spreading out (Figure \ref{alpha0.9_1}c) or collapsing (Figure \ref{alpha0.9_1}d). However, which of these two long-term trends occurs does not depend only on the sign of $\varepsilon$; for example it is not always the case that positive $\varepsilon$ leads to spreading and negative $\varepsilon$ leads to collapse (as in Figures \ref{alpha0.9_1}h and \ref{alpha0.9_1}i). 

Examining these dynamics in more detail we have found that if we consider $\varepsilon$ to be complex, say $\varepsilon=\varepsilon_0 e^{i\theta}$, then for roughly half of the unit circle in terms of $\theta$ we will get spreading and for the other half we get collapse. The two halves will be of approximately equal size and each contiguous, such as $0<\theta<\pi$ and $\pi<\theta<2\pi$, but the exact end points of each interval will depend on the (arbitrary) phase of the eigenvector. 

Panels (e) and (j) in Figure \ref{alpha0.9_1} show the projection of $u(x,t)-u(x,0)\exp(it)$ onto the given eigenvector for positive (blue) and negative (red) perturbations. The overlap of these two projections in the semilogarithmic plot vs. time shown in Panel (e), is a nearly straight line (the black dashed line) up to about $t=12$, and the slope of this line coincides with the eigenvalue $\lambda=0.956$, as expected. Thus, this confirms the exponential
nature of the associated instability with theoretically predicted growth rate.
In Panel (j), the overlap of the two projections indicates two different straight lines (both black dashed lines). At first the negative eigenvalue $\lambda=-0.956$ dominates, as the initial slope corresponds to it, but eventually the positive eigenvalue dominates the motion resulting in a line with the slope of that eigenvalue.

\begin{figure}[h]
\begin{center}
\subfigure[]{\includegraphics[width=4cm]{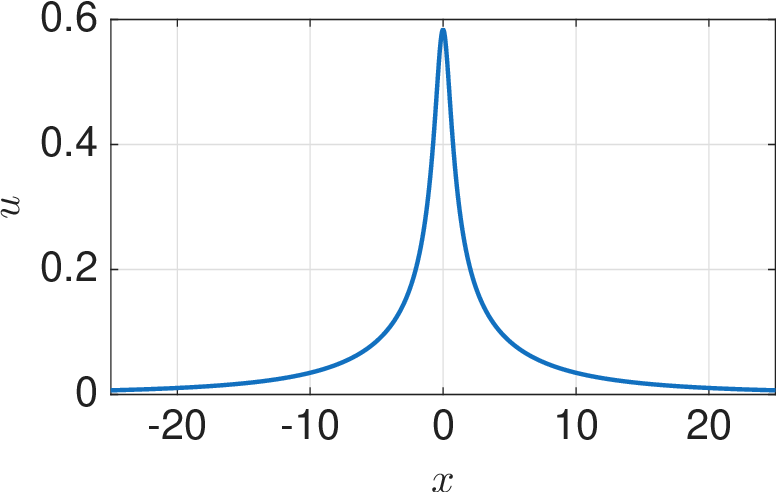}}
\subfigure[]{\includegraphics[width=4cm]
{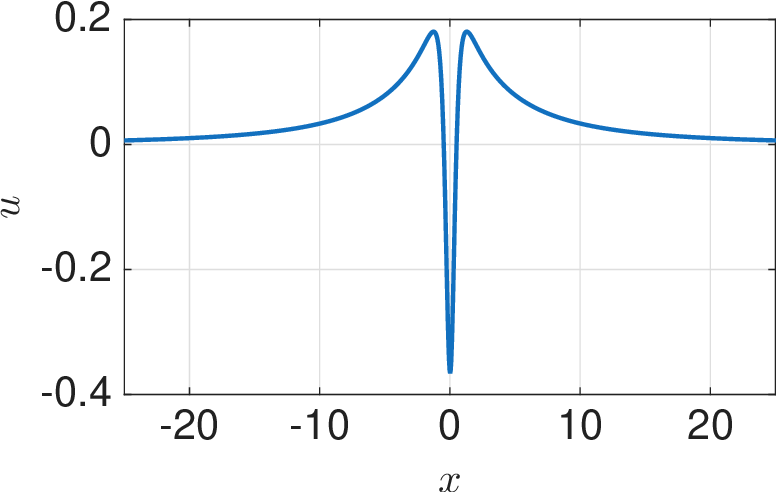}}
\subfigure[]{\includegraphics[width=4cm]{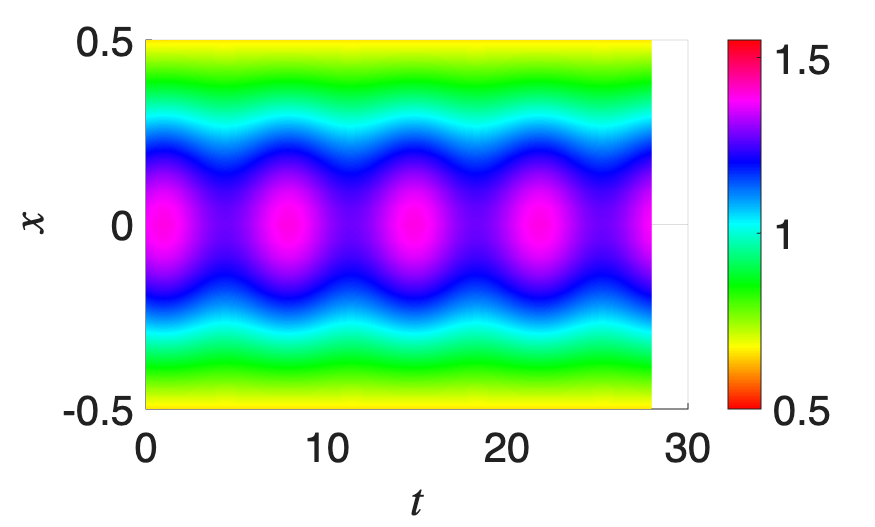}}
\subfigure[]{\includegraphics[width=4cm]
{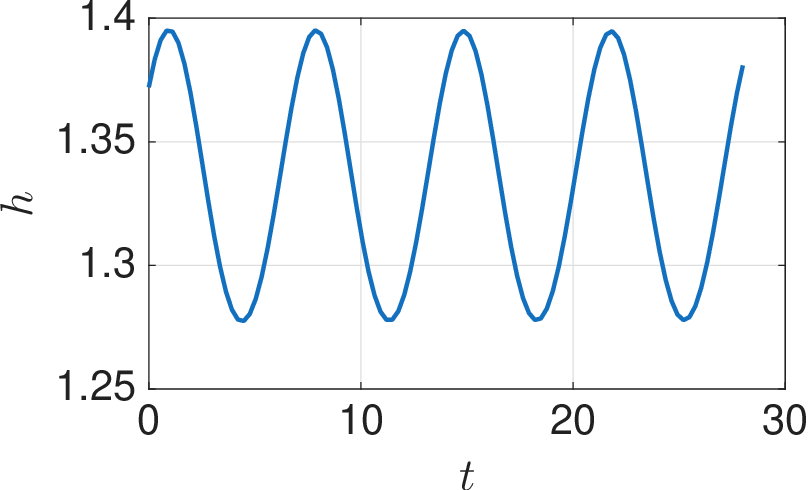}}
\end{center}
\caption{$\alpha=1.3$, single soliton. Panels (a) and (b): real and imaginary parts of eigenvector corresponding to eigenvalue $\lambda=0.8993i$. Panels (c) and (d): contour plot as function of time (c), and height of center point of soliton over time (d). These correspond to a dynamic run using the initial condition `steady state + $\varepsilon \times$eigenvector' where $\varepsilon=0.05i$. Results for the  eigenvalue $\lambda=-0.8993i$, or for other $\varepsilon$ are similar and not shown.}
\label{fig_alpha1.3}
\end{figure}

In Figure \ref{fig_alpha1.3} we summarize the results of a simulation for $\alpha=1.3$, using the eigenvector corresponding to the eigenvalue $\lambda=0.8977 i$. As in Figure~\ref{alpha0.9_1},  the first two panels show the real and imaginary parts of the eigenvectors. The next panel shows a contour plot for a simulation using a perturbation of $\varepsilon=0.05$ and the last one shows the amplitude (height of the soliton at $x=0$) as a function of time. The motion is clearly periodic, and the frequency, as estimated using the local maximum and minimum values from the final panel, corresponds to the imaginary part of the eigenvalue. Indeed, the formerly (i.e., for $\alpha<1$) unstable eigenvalue,
leading to collapse or dispersion in Figure~\ref{alpha0.9_1} leads to a stabilization
and an internal mode oscillation in the case of $\alpha>1$, in line with our
corresponding linearization results.

\section{Two-Soliton Solutions}
\subsection{Bifurcation and stability}
In a way similar to what was illustrated earlier for the kink-bearing problem
in~\cite{DECKER2025100051} (and was proved in~\cite{atanas2}),
due to the oscillations in the tails of single soliton stationary states, 
we expect multiple soliton stationary states to occur for $2<\alpha\le4$.
We consider a two-soliton state, with solitons of equal amplitude, positioned symmetrically with respect to the $y$ axis (we call this case an in-phase two-soliton configuration in the spirit of earlier works such as, e.g.,~\cite{ParkerAcevesPhysD2021Multipulse}). We define the position of each soliton to be the $x$ value of the maximum point of each. In particular, let $x_R$ denote the position of the right soliton and $x_L$ the position of the left soliton. Half of the separation distance $x_R-x_L$ will coincide with the position of the the right soliton $x_R$ and we denote this half-separation distance $\delta$.

\begin{figure}[h]
\begin{center}
\includegraphics[width=10cm]{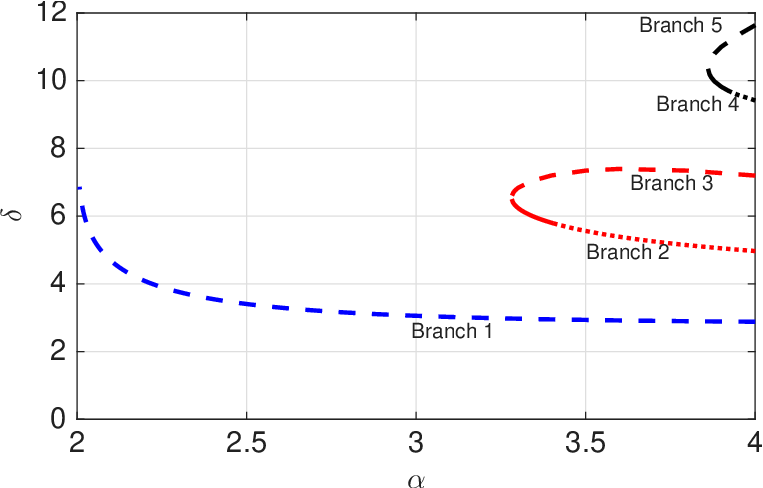} 
\end{center}
\caption{Bifurcation/stability diagram for various two-soliton states. $\alpha$ is the fractional derivative order, and $\delta$ is the half-separation distance (position of the right-side soliton). Dashed and dotted lines indicate instability  and solid lines stability. Dashed lines represent instabilities involving the relative motion of the positions of the solitons, and dotted lines represent instabilities involving changes in the relative amplitudes of the solitons, as will be explained in greater detail later. In the text, the blue dashed line is referred to as branch 1, the lower and upper parts of the red loop as branches 2 and 3 respectively, and the lower and upper parts of the black loop as branches 4 and 5 respectively. }
\label{bifurcate}
\end{figure}

For each value of the exponent $\alpha$ in the interval $2<\alpha \le 4$,
we find that there will be one or more stationary two-soliton states
(or ``solitonic molecules''), each corresponding to a finite value of $\delta$. The $(\alpha , \delta)$ pairs that correspond to stationary states (and their corresponding spectral stability) are shown in Figure \ref{bifurcate}, forming a bifurcation diagram (analogous to the one shown for kinks in~\cite{DECKER2025100051}). Solid lines represent stable states, and both dotted and dashed lines represent unstable states (of different types as explained next).
Indeed, accordingly, there are some significant differences from the corresponding
continuation/bifurcation diagram for kinks, as we highlight below.

\subsection{Short term dynamics and stability}
We now look more closely at the unstable stationary states indicated in Figure  \ref{bifurcate} to determine the nature of the instability. To do that we investigate short-term dynamics (we address longer term dynamics later) corresponding to an initial condition consisting of a stationary state added to a small perturbation multiplied by an eigenvector, as we did in Subsection \ref{sub:dynamics1} for single solitons.

\begin{figure}[h]
\begin{center}
\includegraphics[width=5.7cm]{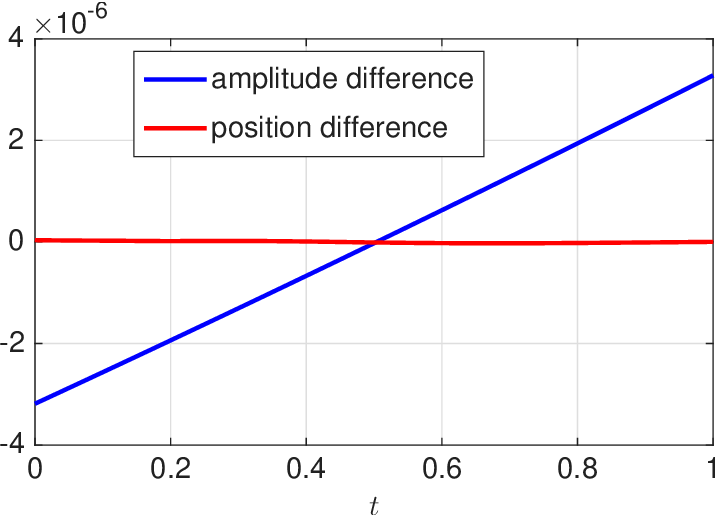} 
\includegraphics[width=5.7cm]{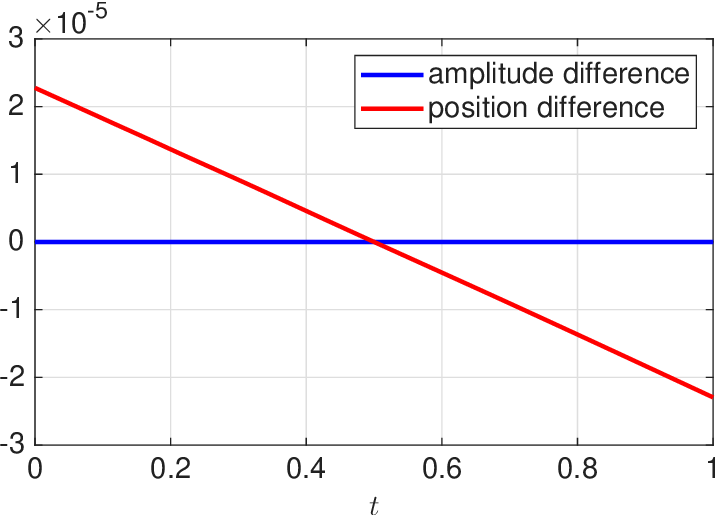} 
\end{center}
\caption{Short term simulation of unstable modes for two-soliton steady states (in phase) with $\alpha=4$, branch 2 (left) and branch 3 (right). For both we measure the change in the difference between soliton positions and the difference between soliton amplitudes. In all cases we use a positive perturbation of $10^{-3}$ multiplied by the eigenvector corresponding to the positive eigenvalue and normalized by subtracting off the mean difference. This indicates that the instability that occurs on part of branch 2 is initially related primarily to the initial relative movement of the amplitudes of the solitons (Type I instability), whereas the instability of all of branch 3 is initially related primarily to the initial relative movement of the positions of the solitons (Type II instability).}
\label{fig:rlud1}
\end{figure}

In Figure \ref{fig:rlud1}, we show the results of running such a simulation for one time unit, where $\alpha=4$ on both branches 2 and 3, using a perturbation of $10^{-3}$ multiplied by the eigenvector corresponding to the positive eigenvalue. In each panel we show the difference in amplitude (blue) and the difference in position (red) for the two solitons, where both have been normalized by subtracting off the mean for each (making it easy to see which changes the most).

Clearly, for branch 2, the perturbation initially affects the amplitude difference more than the position difference, and for branch 3, the reverse is true. This now explains the difference between the dashed and dotted lines in the bifurcation diagram of Figure \ref{bifurcate}, in line also with the statements of~\cite{ParkerAcevesPhysD2021Multipulse}. 
We use dotted lines to represent an instability that involves primarily the relative motion of the amplitudes of the two solitons (i.e., pertaining to symmetry breaking), and dashed lines to represent an instability that involves primarily the relative motion of the positions of the two solitons. These results pertain to in-phase solitons; for out-of-phase solitons see Figures \ref{fig:bifurcateOut} and \ref{fig:outPhase1} in Section \ref{sec:outPhase}.  We remind the reader that in the dark soliton
(real field, also) setting of~\cite{DECKER2025100051}, only the latter destabilization
was available, leading the odd branches to be motionally unstable, while the
even branches were spectrally stable. This leads to the following definition:

\begin{defn}
We define a \textit{Type I instability} as a symmetry-breaking instability, and a \textit{Type II instability} as a relative motion instability. Correspondingly we will refer to the eigenvalue-eigenvector pair that creates a Type I instability as \textit{Type I eigenvalues} and \textit{Type I eigenvectors}, and similarly for \textit{Type II eigenvalues} and \textit{Type II eigenvectors}.
\end{defn}

Going back now to Figure~\ref{bifurcate} and the fundamental (blue) branch,
we find that this smallest distance two-soliton solution is generically unstable (Type II)
for all values of $\alpha$.
In the case of the real field theory of~\cite{DECKER2025100051}, that
is the only instability possible and accordingly, the branches alternate
between (motionally) unstable and stable ones. However, as was illustrated
both theoretically and numerically in the work of~\cite{ParkerAcevesPhysD2021Multipulse},
for the focusing case, this is not so. In particular, the even branches
that were motionally stable, now suffer a Type I instability (for a portion of the branch)
that changes the relative height of the soliton peaks. We now analyze and expand
upon this as a function of $\alpha$. 

\begin{figure}[h]
\begin{center}
\includegraphics[width=8.9cm]{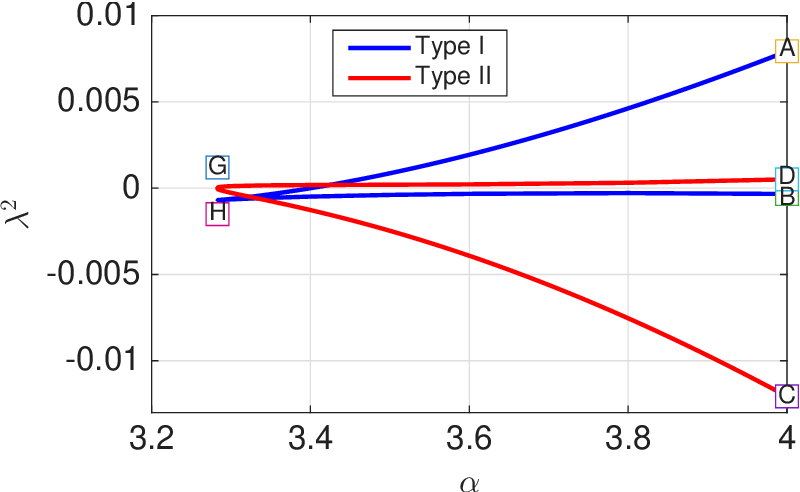} 
\includegraphics[width=8.9cm]{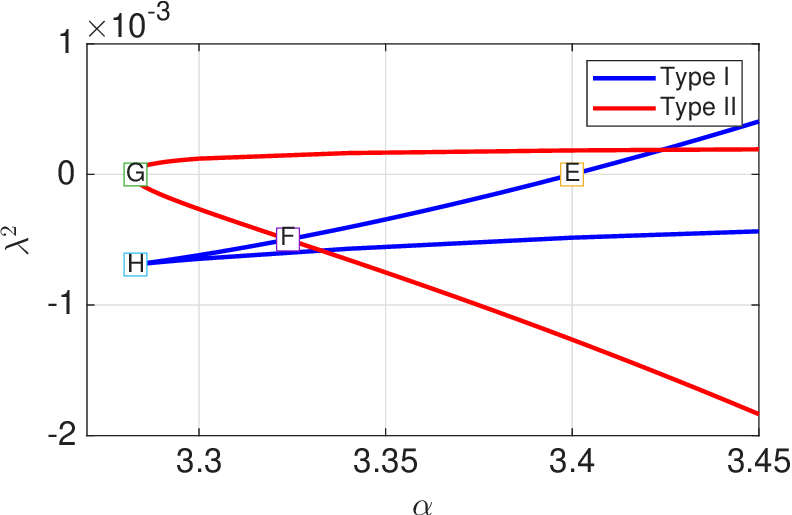} 
\end{center}
\caption{Plots of $\lambda^2$ as a function of $\alpha$ for the red loop in Figure \ref{bifurcate} with the right panel being a zoomed in version of the left panel. Points A and C correspond to the the point on the lower branch of the red loop at $\alpha=4$ in Figure \ref{bifurcate}, and B and D to the upper branch at $\alpha=4$. Points E and G are where real eigenvalues become imaginary or vice versa (paths cross the horizontal axis). 
In the first, the Type I symmetry breaking instability disappears, while as we cross the
second, the positional variation (as a Type II instability) becomes a destabilized 
eigendirection.}
\label{movingEigsSq_in}
\end{figure}

In Figure \ref{movingEigsSq_in} we show graphs of $\lambda^2$ as a function of $\alpha$ (similar to Figure \ref{alphaSquared} for single solitons) for branch 2 and branch 3, i.e., the red loop in Fig.~\ref{bifurcate}. The right panel is a zoomed-in version of the left one. Note that positive values of $\lambda^2$ correspond to real 
eigenvalues $\lambda$ and negative values of $\lambda^2$ correspond
to imaginary $\lambda$ (hence, a breathing eigenmode). 
The blue path AHB traces the motion of the Type I eigenvalues on branches
2 and 3, starting at A (lower branch of red loop at 
$\alpha=4$) and ending at B (upper branch of red loop at $\alpha=4$) in Figure \ref{bifurcate}. 
Similarly the red path CGD traces the motion of the Type II eigenvalues, starting at C and ending at D, the lower and upper branches of the red loop at $\alpha=4$. Point E 
($\alpha \approx 3.4$, lower branch of red loop) is where the Type I eigenvalues switch to the imaginary axis, rendering for the first time (in this model, to our knowledge)
these multi-pulses linearly stable. This means that the $\alpha$ value of point E is also the $\alpha$ in Figure \ref{bifurcate} where the dotted line changes to a solid line (first bifurcation: 
a pitchfork associated with the symmetry breaking), and the $\alpha$ value of point G 
is where the solid line changes to a dashed line (second bifurcation: a saddle-center
one). The paths AH and CG correspond to the lower part of that loop, whereas HB and GD correspond to the upper part. Point F is where the two pairs of eigenvalues (Type I and Type II) cross each other on the imaginary
axis; it is not a bifurcation point for in-phase solitons.

\begin{figure}[h]
\begin{center} 
\includegraphics[width=14cm]
{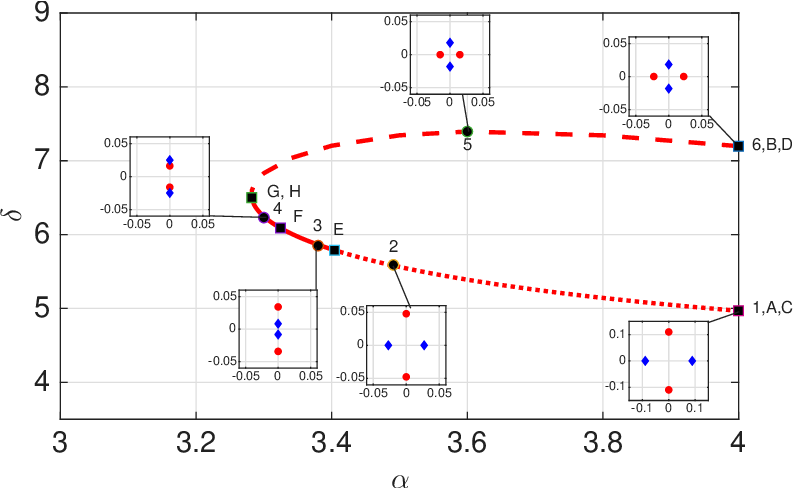} 
\end{center}
\caption{Bifurcation diagram of branches 2 and 3 from Figure \ref{bifurcate} (red loop only) with spectral plots and key labeled points from Figure \ref{movingEigsSq_in} arranged with respect to each other on the loop. See Table \ref{tab:alphaValues} for values of $\alpha$ for all labeled points. The Type I eigenvalues (blue diamonds) correspond to the blue curves in Figure \ref{movingEigsSq_in}, and similarly for the Type  II eigenvalues (red circles) and red curves. Note: \textit{spectral plots do not include} the two pairs of eigenvalues within $10^{-5}$ of the origin.}
\label{fig:annotatedBifurcationIn}
\end{figure}

We now present a plot in Figure \ref{fig:annotatedBifurcationIn}, where we combine information from Figures \ref{bifurcate} and \ref{movingEigsSq_in}. This figure shows the red loop from \ref{bifurcate} along with the labeled points from Figure \ref{movingEigsSq_in} positioned along the loop at their respective $(\alpha,\delta)$ values, as well as spectral plots showing how the Type I and Type II eigenvalues move in the spectral plane as the loop is traversed. These spectral plots can be identified with the numerals 1-6. The numerals label the circular markers at the $(\alpha,\delta)$ values along the red loop used to produce the corresponding plots (except at $\alpha=4$  where the square and circular markers coincide so we use only the square marker). See Table \ref{tab:alphaValues} for the $\alpha$ values of all labeled points. We now traverse the red loop and describe what happens as we move through the square and circular markers.

For all spectral plots there are four isolated eigenvalues (two opposite pairs).  Two pairs of eigenvalues, within $10^{-5}$ of the origin \textit{are not shown} as our focus is on the Type I and Type II eigenvalues. We start with plot 1 at $\alpha=4$, 
associated with branch 2, indicating an instability of Type  (blue diamond eigenvalues are real). This is the same
instability that was identified in~\cite{ParkerAcevesPhysD2021Multipulse} for
this branch. In plot 2, $\alpha$ has decreased, but is still in the dotted (red) line region of the figure; i.e., it is still unstable. The real-valued Type I eigenvalues (blue diamonds) are, however, moving toward the origin. In plot 3, $\alpha$ has decreased further and the Type I eigenvalues have passed through point E (bifurcation point) into the solid red region of stability. 

This is
the first {\it stabilization} bifurcation; it is a symmetry restoring that
is {\it not possible} for $\alpha=4$, however it becomes {\it only possible
for the fractional values of $\alpha$} considered herein. This is a fundamental
finding of the present work, to be contrasted with the {\it generic instability}
of two-soliton solutions found in~\cite{ParkerAcevesPhysD2021Multipulse}.
Consequently the Type I eigenvalues have passed through the origin and are now heading upwards on the imaginary axis, while another pair of isolated imaginary eigenvalues (Type II)
is moving toward the origin. 
In plot 4 we are still in the solid-red line region and the two pairs of isolated imaginary eigenvalues have passed each other. Thus the Type II eigenvalues associated with the
translation of the solitary waves (and relevant oscillation) have a lower frequency,
while the Type I eigenvalues asymmetrizing the amplitudes correspond to a higher frequency.
In plot 5 we have moved onto branch 3, the dotted red line (Type II instability); this is the second bifurcation, a saddle-center bifurcation. 
Here, the destabilization of the branch takes place through motion of the
centers
and the system wants to depart from this saddle configuration (which in the
energy landscape as a function of the pulse relative 
position corresponds to an energy maximum).
In plot 6, we have come back to $\alpha=4$, now on branch 3, still unstable
through the pulse center motion. 

Summarizing the motion of the isolated eigenvalues, we can say that the initially real Type I pair moves toward the origin, switches to the imaginary axis (stabilizing the system against symmetry breaking), then passes through the initially imaginary Type II pair, after which the Type II pair switches to the real axis and destabilizes
the configuration anew, as was predicted for $\alpha=4$ in~\cite{ParkerAcevesPhysD2021Multipulse}.  
This complete eigenvalue trajectory reconciles the stability
findings of~\cite{ParkerAcevesPhysD2021Multipulse} with the
fractional model stabilization conclusions presented herein.

\begin{table}[h]
    \centering
    \begin{tabular}{c|c|c|c|c|c|c|c|c|c|c|c|c|c|c}
       point & 1 & A  & C & 2 & E & 3 & F & 4 & G & H & 5 & 6 &B & D\\
       \hline
       $\alpha$ & 4 & 4 & 4 & 3.49 & 3.4 & 3.39 & 3.32 & 3.3 & 3.28 & 3.28 & 3.6 & 4 & 4 & 4\\
    \end{tabular}
    \caption{Alpha values of labeled points in Figures \ref{movingEigsSq_in} and \ref{fig:annotatedBifurcationIn}}.
    \label{tab:alphaValues}
\end{table}

\subsection{Longer term two-soliton dynamics}

\begin{figure}[h]
\begin{center}
\subfigure[]{\includegraphics[width=4.5cm]
{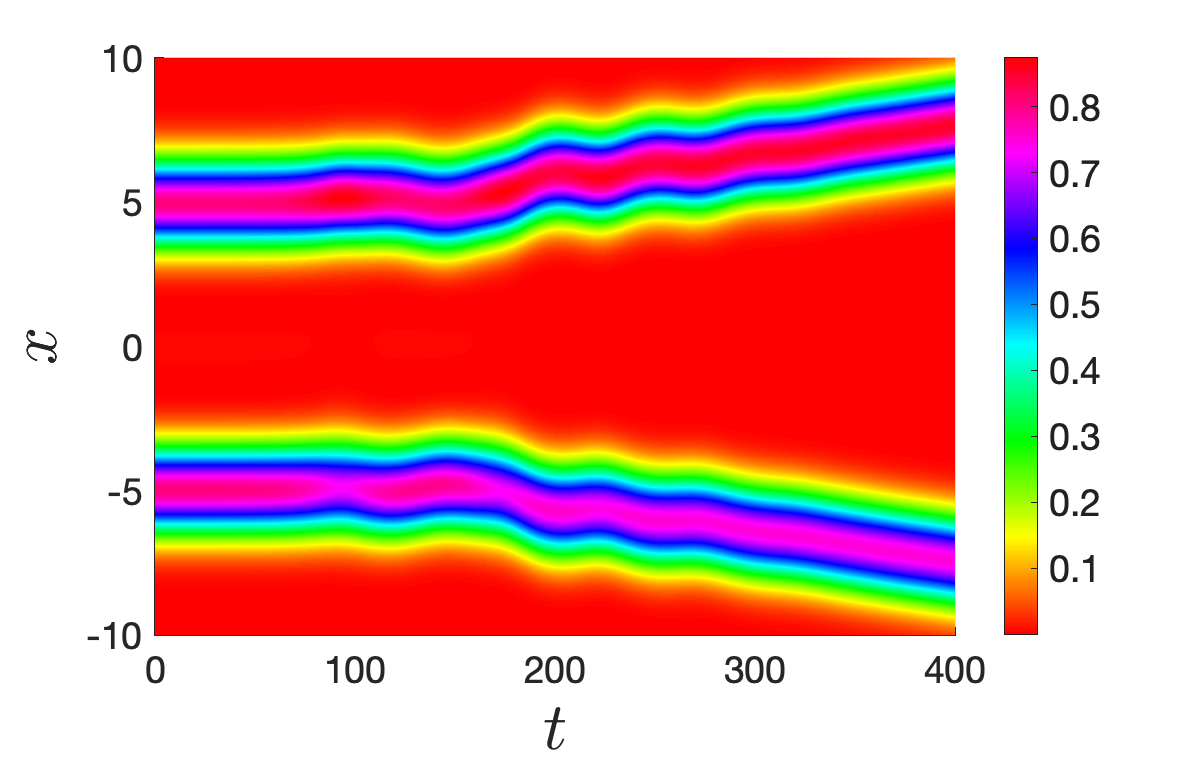} }
\subfigure[]{\includegraphics[width=4.5cm]
{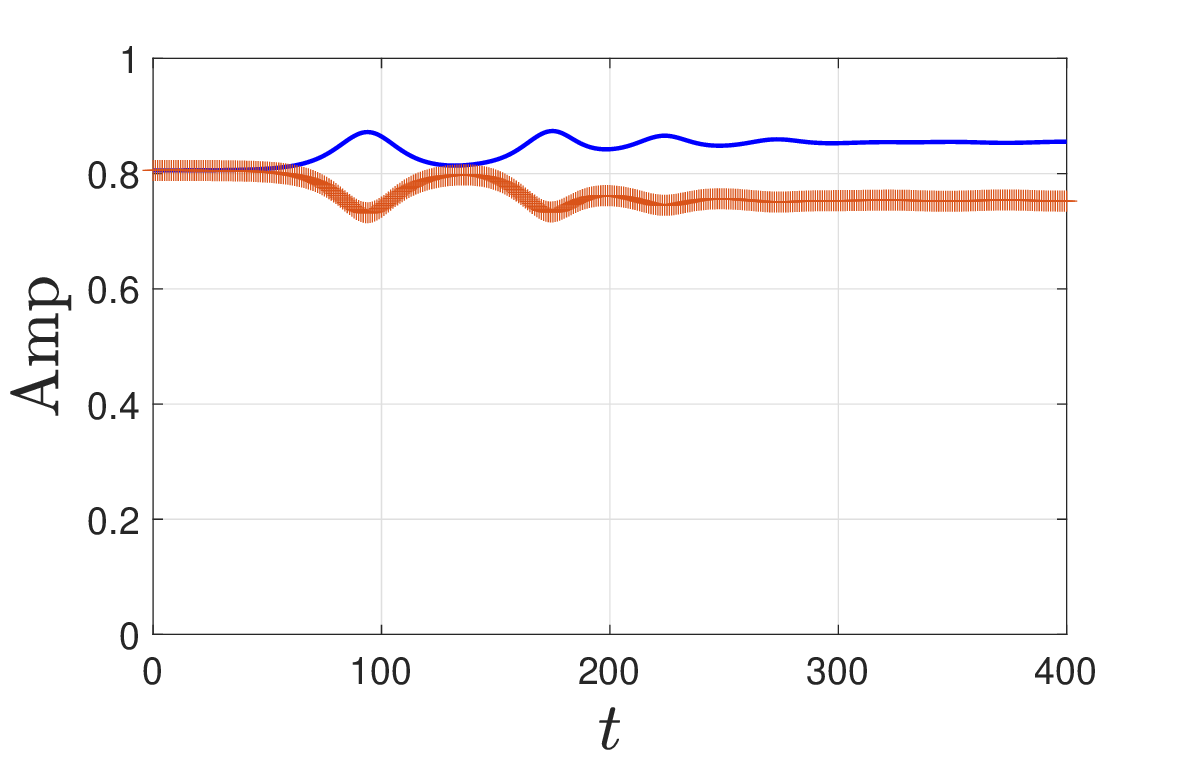} }
\subfigure[]{\includegraphics[width=4.5cm]
{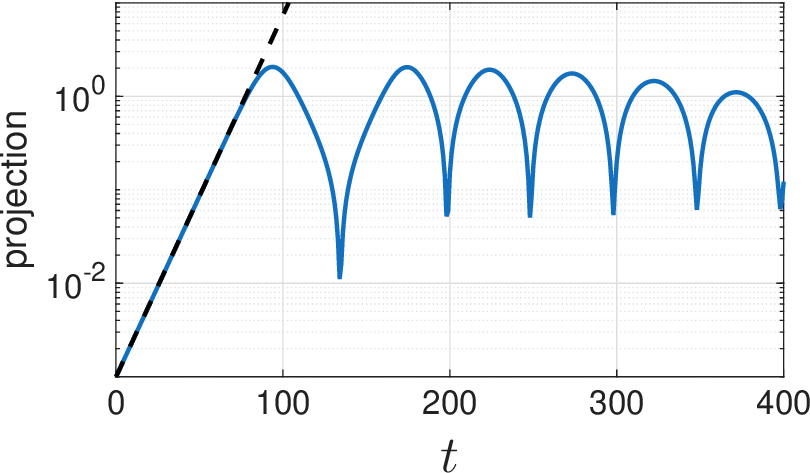}} 
\end{center}
\caption{Long term simulation, branch 2 (unstable part), positive perturbation ($10^{-3})$ along
the growing eigendirection, using $\alpha=4$ as an example. Panel (a): surface plot. Panel (b): both amplitudes (right soliton in solid blue, left soliton in red with '+' markers). The slight difference in amplitude between the upper and lower paths in the surface plot  in (a) (shown by the coloring and the path width) is more clearly demonstrated in the amplitude plot in (b). The destabilization due to symmetry breaking in (b) is eventually manifested
by the pulse separation in (a).
 Panel (c): projection of the motion onto the eigenvector corresponding to the positive eigenvalue $\lambda=0.089$ (positive and negative perturbations are the same). The initial slope of the projection is $0.089$, giving an accurate growth rate.}
\label{fig:longDynamics1}
\end{figure}

\begin{figure}[h]
\begin{center}
\subfigure[]{\includegraphics[width=3.4cm]
{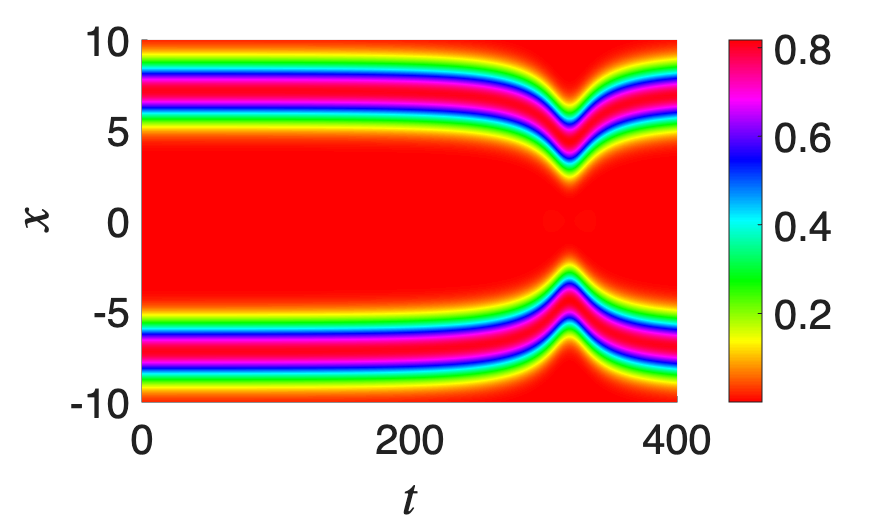} }
\subfigure[]{\includegraphics[width=3.4cm]
{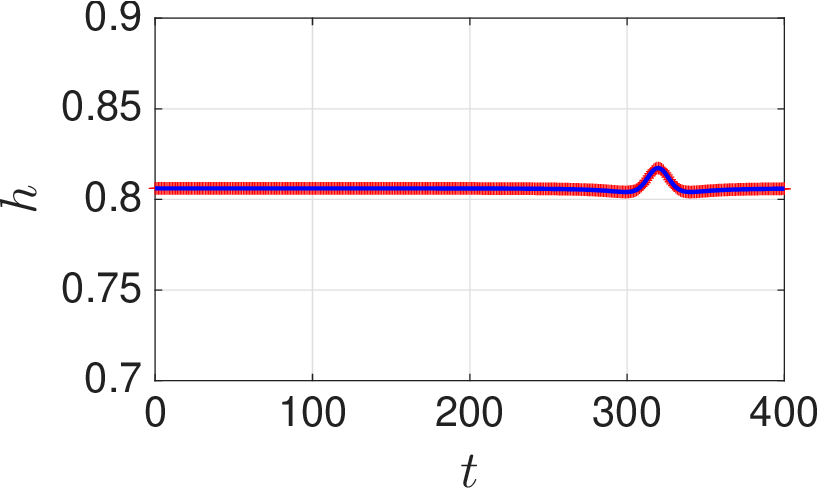} }
\subfigure[]{\includegraphics[width=3.4cm]
{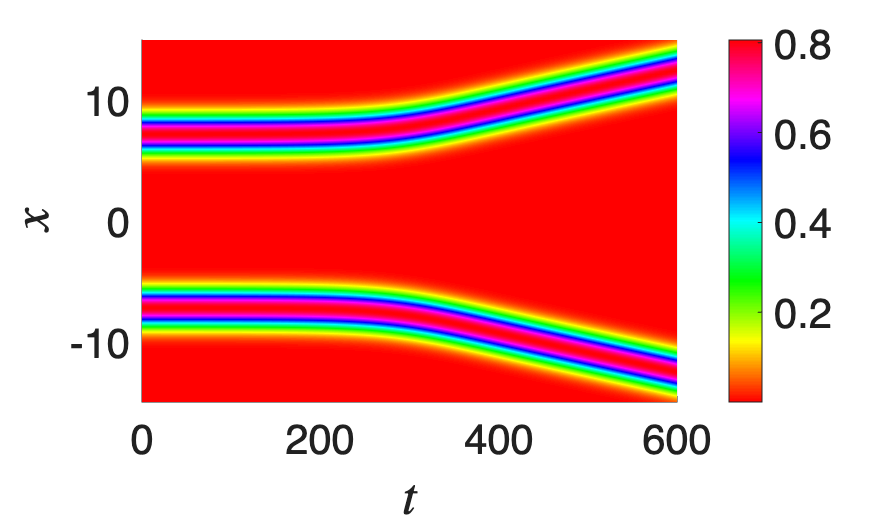} }
\subfigure[]{\includegraphics[width=3.4cm]
{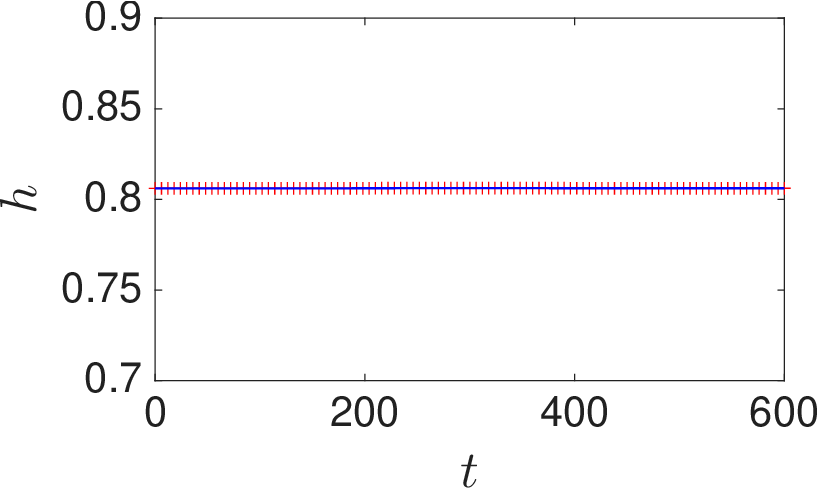} }
\subfigure[]{\includegraphics[width=3.4cm]
{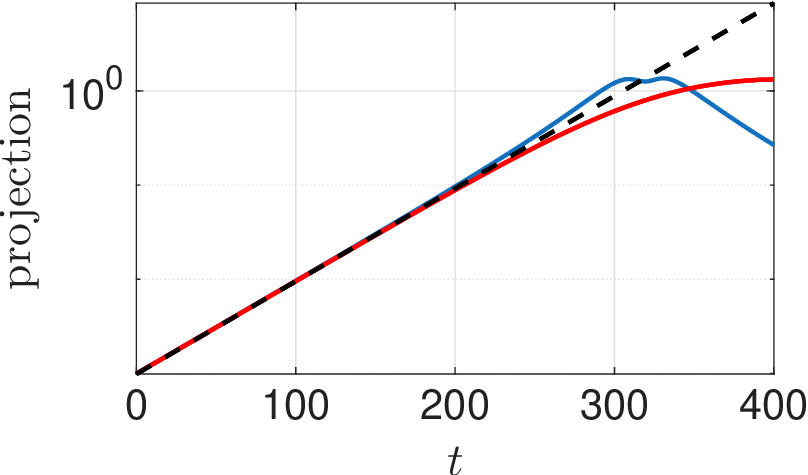}} 
\end{center}
\caption{Long term perturbed dynamics for branch 3 (unstable), with perturbation corresponding to the eigenvector associated with the positive real eigenvalue. Positive perturbation (panels (a) and (b)) and negative perturbation (panels (b) and (c)). Surface plots are followed by plots of both amplitudes (in these cases the right and left amplitudes coincide). The final plot represents the projection onto an eigenvector (for both positive and negative perturbations); initial slope where both projections coincide is $0.226$ and the eigenvalue is $\lambda=0.226$, verifying the growth rate of the projection.}
\label{fig:longDynamics2}
\end{figure}

For longer-term dynamics of unstable, two-soliton settings see Figure \ref{fig:longDynamics1} for branch 2 dynamics and Figure \ref{fig:longDynamics2} for branch 3 dynamics. In Figure \ref{fig:longDynamics1}, we see that though Figure \ref{fig:rlud1} indicates that initially the difference in amplitude of the two solitons is excited to a much greater extent than the difference in position, for this longer-term simulation the amplitude mode begins to interact with the position mode ---with the two being coupled by nonlinearity---
so that by about $t=150$ the soliton positions begin to separate. The separation rate of the positions becomes constant and the two amplitudes stabilize at a fixed distance apart, as the pulses are sufficiently well separated that they
do not interact any further. A projection onto the relevant eigenvector shows that the initial growth rate corresponds to the value of the positive eigenvalue $\lambda=0.0891$,
until eventually (per the separation of the solitary waves), the relevant projection
is saturated and subsequently oscillates. 
When a negative perturbation is used, the surface plot in the first panel 
of Figure \ref{fig:longDynamics1}
is reflected about the $x$-axis, and the second plot is reflected about its axis of symmetry (near amplitude of $0.8$).

In Figure \ref{fig:longDynamics2} (the branch 3 simulation) we see that the dynamics predicted by the short-term results in Figure \ref{fig:rlud1}
is more closely followed in the long term than was the case for branch 2 (Figure \ref{fig:longDynamics1}). In particular we see that the positions of the two solitons either approach each other (resulting in a bounce) or they depart from each other at a constant rate, depending on whether a positive or negative perturbation is used. As expected a projection onto eigenvector plot shows that the growth rate of the projection matches the positive eigenvalue. In either case, the parity of the initial 
evolution is definite, rendering it thus far less likely for this translational
mode to couple to the mode leading to amplitude variation.

\subsection{Two-soliton out-of-phase results}
\label{sec:outPhase}

\begin{figure}[h]
\begin{center}
\includegraphics[width=10.0cm]{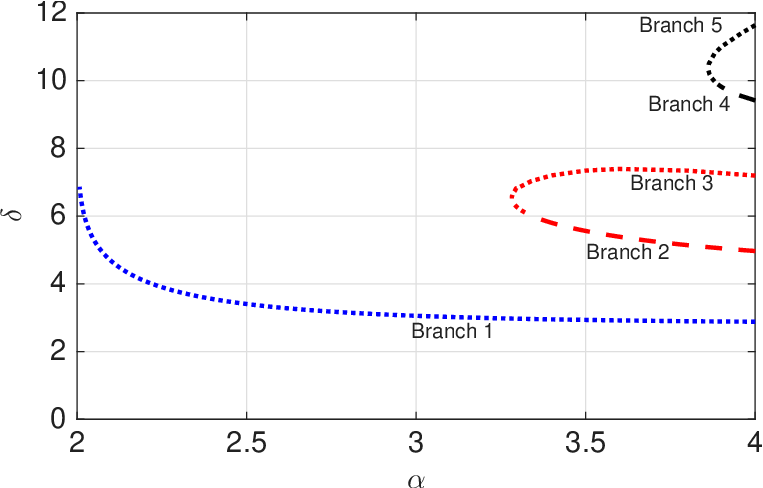} 
\end{center}
\caption{Bifurcation diagram for out-of-phase solitons. Dotted branches reflect Type I
symmetry-breaking instabilities, while dashed ones, Type II instabilities related to 
translational motion.}
\label{fig:bifurcateOut}
\end{figure}

In line with the earlier investigation of~\cite{ParkerAcevesPhysD2021Multipulse},
we also explore the family of out-of-phase bright solitary pulses.
Such out-of-phase states consist of one soliton with positive amplitude and another with negative amplitude. The locations of stationary states (in terms of $(\alpha,\delta$) space) do not change from Figure \ref{bifurcate}.  However, the stability type does change; see Figure \ref{fig:bifurcateOut}. 

\begin{figure}[h]
\begin{center}
\includegraphics[width=8.9cm]{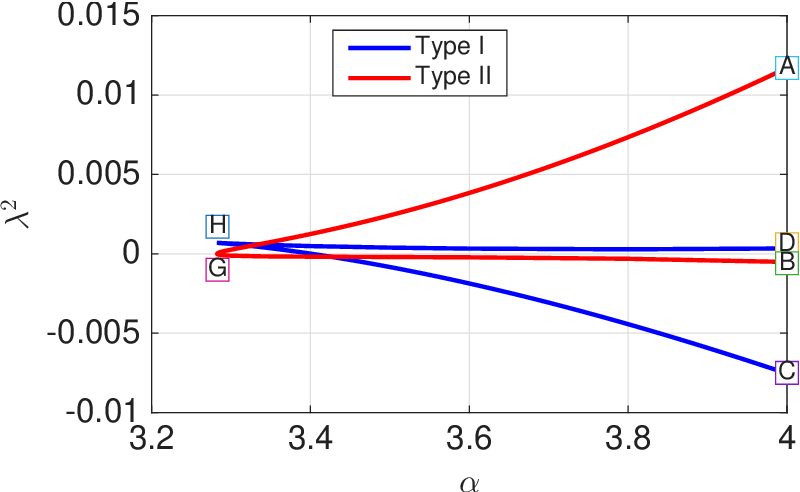} 
\includegraphics[width=8.9cm]{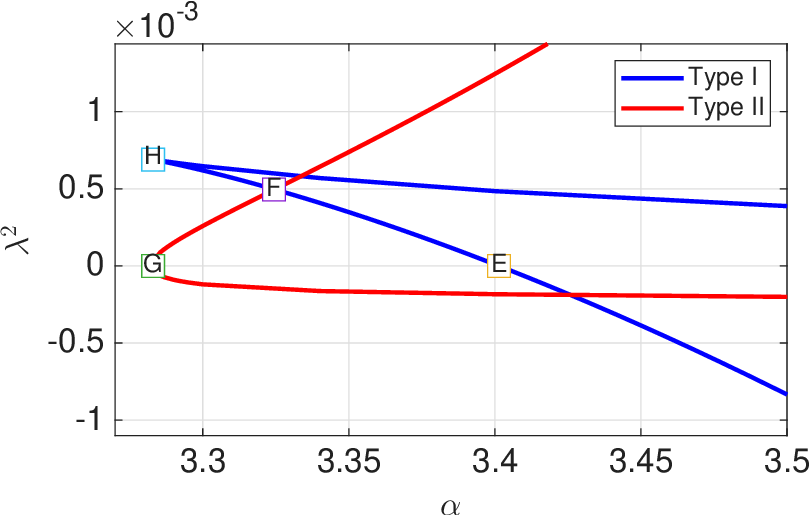} 
\end{center}
\caption{Out-of-phase plots of $\lambda^2$ as a function of $\alpha$ for the red loop in Figure \ref{fig:bifurcateOut}, similar to Figure \ref{movingEigsSq_in} for in-phase solitons.}
\label{movingEigsSq_out}
\end{figure}

For branch 1 (blue dotted) the stability as changed from Type II to Type I. For branches 2 and 3, as we did with the in-phase case, we can traverse the red loop in Figure \ref{fig:bifurcateOut} in a clockwise direction and can track the motion of the four isolated eigenvalues (excluding the quartet of eigenvalues within $10^{-5}$ of the origin). Again we find that we begin with a pair of real eigenvalues and a pair of imaginary ones, and end with a similar configuration, but as opposed to the in-phase case, this time the real eigenvalues at $\alpha=4$ on branch 2 are of Type II, and the real eigenvalues at $\alpha=4$, branch 3, are of Type I. Furthermore, this time the pairs pass through each other on the real axis rather than on the imaginary axis. In Figure \ref{movingEigsSq_out} we show the graph of $\lambda^2$ as a function of $\alpha$, as in Figure \ref{movingEigsSq_in} for in-phase solitons.

Of note when comparing the in-phase case (Figure \ref{movingEigsSq_in}) to the out-of-phase case (Figure \ref{movingEigsSq_out}) is that they are reflections of each other with respect to the horizontal axis. 
This implies that when the translational modes (odd branches) are unstable
in the in-phase case, they are associated with imaginary eigenvalues in the
out-of-phase case, while the symmetry-breaking even branches do not feature
out-of-phase destabilizing symmetry-breaking. Instead the latter are translationally
unstable, while the former (odd branches) feature symmetry-breaking.
This is reflected in the interchange of eigenvalues that occurs on the real axis (positive $\lambda^2$).  The interchange on the real axis has a further consequence; point F in Figure  \ref{movingEigsSq_out}, where the interchange occurs, becomes the bifurcation point in Figure \ref{fig:bifurcateOut}, where the instability type changes from being position related to being amplitude related (dashed to dotted in the figure). Point E is no longer a bifurcation point as it was for in-phase solitons.  
Indeed, here, contrary to the in-phase case, rather than complete stabilization,
the branch 2 will feature a ``double destabilization'' near its turning
point, due to both translational and symmetry-breaking instabilities.

\begin{figure}[h]
\begin{center}
\includegraphics[width=6.0cm,height=4cm]{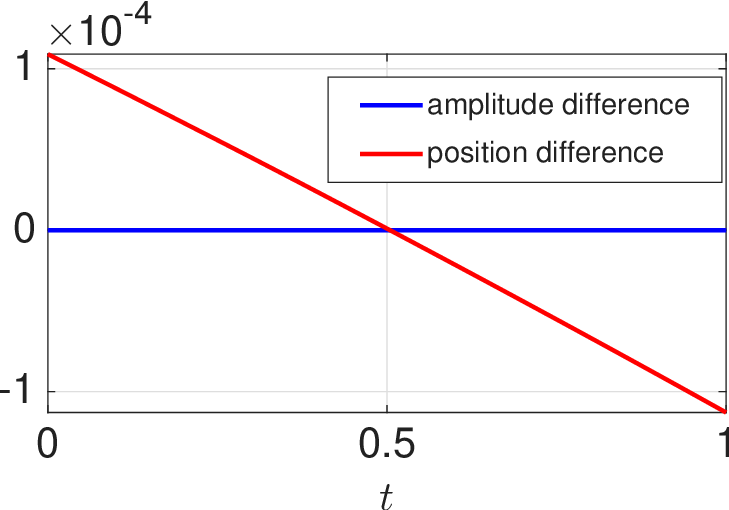} 
\includegraphics[width=6.0cm,height=4cm]{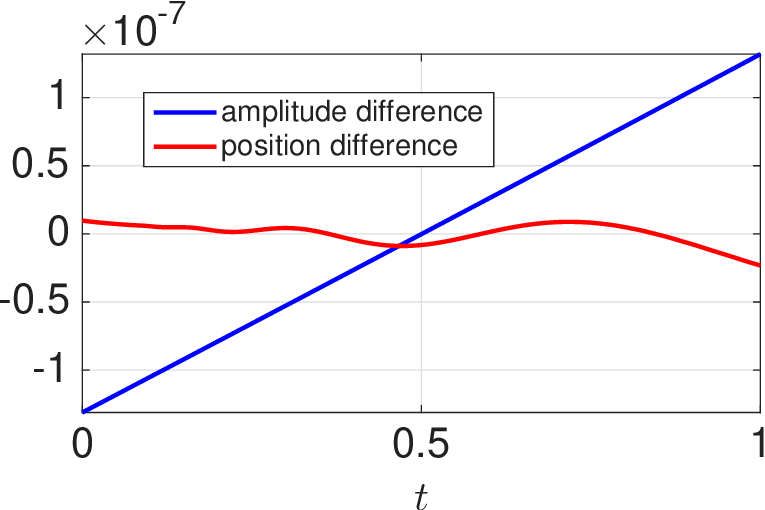} 
\end{center}
\caption{Unstable modes for two soliton steady states (out-of-phase) with $\alpha=4$, branch 2 (left) and branch 3 (right), as in Figure \ref{fig:rlud1} for in-phase two soliton states. The perturbation amplitude is $10^{-3}$; for an equivalent negative perturbation the image in the left panel is the mirror image with respect to the $x$-axis and in the right panel the blue line is reflected about the $x$-axis and the red curve is the same.}
\label{fig:outPhase1}
\end{figure}

See Figure \ref{fig:outPhase1} for short term dynamics in the out-of-phase case. As shown in the bifurcation diagram in Figure \ref{fig:bifurcateOut}, we see that at $\alpha=4$ the results have been reversed from the in-phase case; at the beginning of the red loop (branch 2, $\alpha=4$) the instability is of the position varying Type II, while at the end (branch 3, $\alpha=4$) the amplitude difference (Type II instability) dominates (though somewhat less dramatically,
and the position difference is mildly excited). These behaviors are continued up to the bifurcation point (where dotted changes to dashed in Figure \ref{fig:bifurcateOut}) from both sides. Once again, the saddle-center bifurcation concerns the position dynamics
with a local positional minimizer and a local saddle point colliding and
disappearing.

\begin{figure}[h]
\begin{center}
\subfigure[]{\includegraphics[width=4.5cm,height=3.1cm]{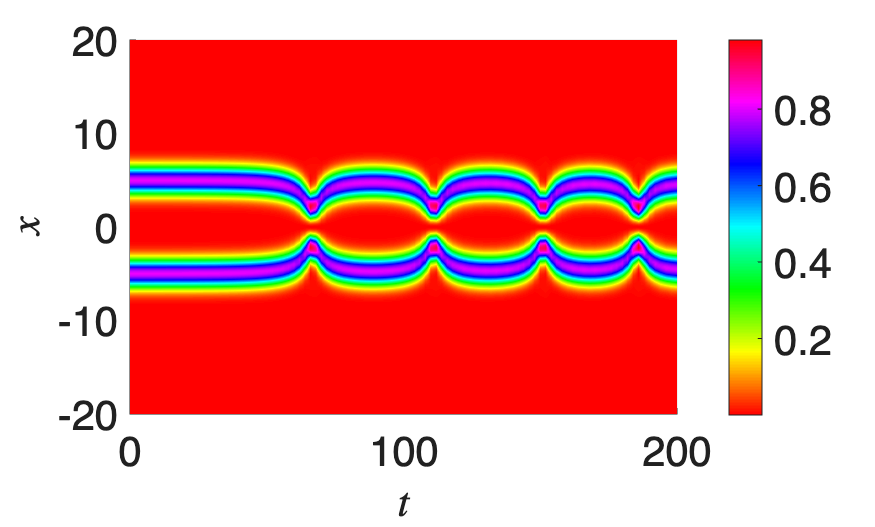} }
\subfigure[]{\includegraphics[width=4.5cm,height=3.1cm]{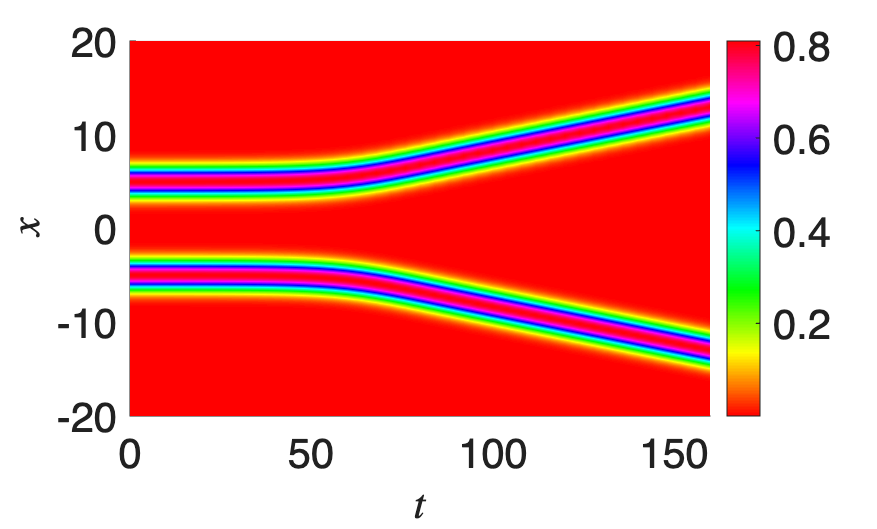}} 
\subfigure[]{\includegraphics[width=4.5cm,height=3.1cm]{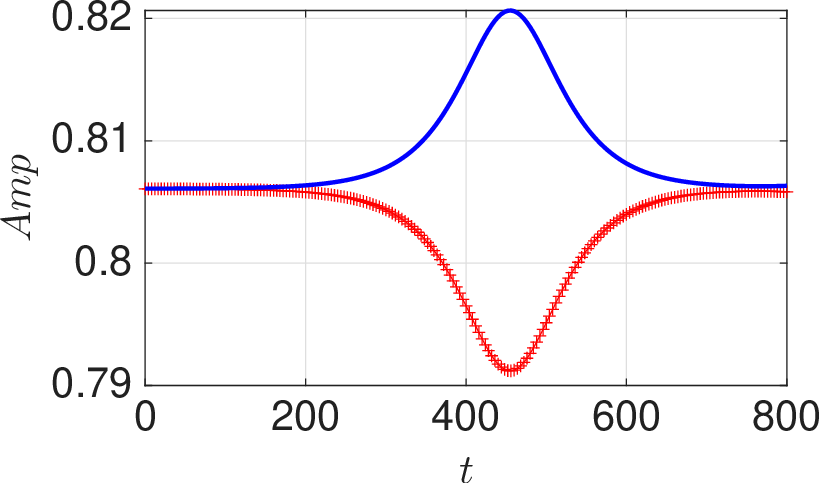} }
\end{center}
\caption{Long term perturbed dynamics for out-of-phase solitons, $\alpha=4$. Panels (a) and (b) are surface plots from branch 2 (positive and negative perturbations, respectively), reflecting translationally unstable perturbations leading the
pulses to approach or separate. Panel (c) plots both the right soliton amplitude (blue line) and the left soliton amplitude (red crosses) from branch 3, for a 
symmetry-breaking (Type I) perturbation.}
\label{fig:outPhase2}
\end{figure}

These short-term results have, as usual, a bearing towards the longer-term results of Figure \ref{fig:outPhase2}. There we see that for most of branch 2 (prior to the bifurcation point) the solitons either separate or approach each other and bounce, depending on the sign of the perturbation. 
In this case, the parity preservation (up to roundoff error) prevents the
``contamination'' of the mode involving the symmetry breaking.
On the other hand, for branch 3 (and its continuation toward branch 2 up to the bifurcation points) the positions do not change much (not shown) and the amplitude difference grows for some time, before being reduced to a value close to the starting value (after which the behavior is repeated---but not periodically---again not shown,
as the latter evolves over much longer time scales).

In the above analysis, we have been mostly focused on the red loop (i.e. branches 2 and 3) of Figures \ref{bifurcate} and \ref{fig:bifurcateOut}. As shown in those figures, branch 1 has an instability of translational type for in-phase solitons, and of the 
symmetry-breaking type for out-of-phase solitons. The black loop, consisting of branches 4 and 5, demonstrates behaviors very similar to those of the red loop. However, in the case of initial excitation of the symmetry-breaking mode, the translational mode is also excited sooner and to a greater degree than for the red loop cases.

\section{Conclusions and Future Challenges}

In the present work we have revisited from a numerical perspective the 
existence, stability and dynamical problem of single- and two-solitary
waves in the context of the fractional focusing NLS equation.
For the single solitary wave, we identified the regime of stability
$\alpha>d$, where $\alpha$ is the fractional exponent and $d$ the operator
dimensionality. We have showcased the instability as pertaining to
an eigenvalue pair that is real for $\alpha<d$ and imaginary
(i.e., associated with an internal mode) for $\alpha>d$. 
We also visualized the interval of $\alpha$ where the pulses are
monotonic in their tails (i.e., $\alpha<2$) and showed their increasing degree
of non-monotonicity when $2 < \alpha < 4$. Turning then to
two-soliton solutions, we found that stationary states do not
exist for $\alpha<2$, but arise for arbitrary $\alpha$'s such 
that $2< \alpha < 4$. As $\alpha$ is increased toward $4$, the
number of such branches increases (in pairs, through saddle-center
bifurcations), becoming infinite as $\alpha \rightarrow 4$. 
While our results for $\alpha=4$ connected with earlier
findings suggesting instability of all of these branches
(odd ones due to translations, and even ones due to symmetry breaking),
a remarkable finding is that for an interval of non-integer $\alpha$'s,
in-phase two-soliton pairs become stabilized. Remarkably, and given that
the picture is flipped (between translations and symmetry-breakings) for
out-of-phase states, the latter can never be spectrally stable. We also
explored the implications of destabilization for single solitons
(collapse/dispersion) and multisolitons (where the pulses are either
set to motion, or led to asymmetrize their amplitudes).

Naturally, this investigation poses an array of interesting problems
for future study. For a single solitary wave, devising an effective
internal degree of freedom that could capture the oscillation (for $\alpha>d$)
and destabilization toward collapse/dispersion (for $\alpha<d$) would naturally
be of interest. In the context of multiple solitary waves, an aspect
that is sorely missing involves an effective description of the 
soliton-soliton interaction in this fractional setting, where power law
effective dynamics is anticipated. While similar steps have been taken in 
the context of long-range interacting Klein-Gordon kinks~\cite{Manton_2019,christov1},
we are not aware of ones such for the fractional focusing NLS setting.
At the reduced level description, our analysis suggested that in the 2
solitary wave setting, a 4-dimensional description associated with the
relative position of the pulses, and their relative amplitude (describing the
symmetry breaking) would be particularly suitable. While these modes 
both influence stability and do not ``cross talk'' at the linear stability
level, we have made the case for their interaction via nonlinearity
and their driving of the full pulse dynamics within the system of interest.
In-phase and out-of-phase cases are both worth exploring and feature
a symmetry between their respective descriptions. Finally, a natural
further step to the ones above concerns the extensions of such considerations
in higher-dimensional settings; see, e.g., at the real field-theoretic
level motivating the work of~\cite{jin2025effectdiscretenessdomainwall}.

\section*{Acknowledgments}
This material is partially based upon work supported by the U.S. National Science Foundation under the award PHY-2408988 (PGK). This research was partly conducted while P.G.K. was 
visiting the Okinawa Institute of Science and
Technology (OIST) through the Theoretical Sciences Visiting Program (TSVP)
and the Sydney Mathematical Research Institute (SMRI) under an SMRI Visiting
Fellowship. 
This work was also supported by the Simons Foundation
[SFI-MPS-SFM-00011048, P.G.K.]. The authors thank Prof. M. de Sterke
of the University of Sydney for useful discussions on the subject of
fractional dispersion and its experimental realization.

\bibliographystyle{apsrev4-1}
 \bibliography{Bibfile3}

\end{document}